\documentclass[journal=jacsat,manuscript=article]{achemso}
\setkeys{acs}{maxauthors=10,etalmode=truncate,chaptertitle =true,articletitle=true}

\usepackage{amssymb}
\usepackage{amsmath}
\usepackage[usenames,dvipsnames]{xcolor}
\usepackage{graphicx}
\usepackage{float}
\usepackage[normalem]{ulem}
\usepackage{caption}
\usepackage{subcaption}
\usepackage{natbib}
\usepackage{xcolor}

\usepackage{xcolor}

\usepackage{multirow}
\usepackage{bm}
\usepackage[version=4]{mhchem} 

\SectionNumbersOn

\title{Prediction of Photodynamics of 200 nm Excited Cyclobutanone with Linear Response Electronic Structure and Ab Initio Multiple Spawning}
\author{Diptarka Hait}
\affiliation
{{Department of Chemistry and The PULSE Institute, Stanford University, Stanford, California 94305, United States}}
\alsoaffiliation{SLAC National Accelerator Laboratory, Menlo Park, California 94024, United States}
\altaffiliation{These authors contributed equally to this work.}
\author{Dean Lahana}
\affiliation
{{Department of Chemistry and The PULSE Institute, Stanford University, Stanford, California 94305, United States}}
\alsoaffiliation{SLAC National Accelerator Laboratory, Menlo Park, California 94024, United States}
\altaffiliation{These authors contributed equally to this work.}
\author{O. Jonathan Fajen}
\affiliation
{{Department of Chemistry and The PULSE Institute, Stanford University, Stanford, California 94305, United States}}
\alsoaffiliation{SLAC National Accelerator Laboratory, Menlo Park, California 94024, United States}
\altaffiliation{These authors contributed equally to this work.}
\author{Amiel S. P. Paz}
\affiliation
{{Department of Chemistry and The PULSE Institute, Stanford University, Stanford, California 94305, United States}}
\alsoaffiliation{SLAC National Accelerator Laboratory, Menlo Park, California 94024, United States}
\author{Pablo A. Unzueta}
\affiliation
{{Department of Chemistry and The PULSE Institute, Stanford University, Stanford, California 94305, United States}}
\alsoaffiliation{SLAC National Accelerator Laboratory, Menlo Park, California 94024, United States}
\author{Bhaskar Rana}
\affiliation
{{Department of Chemistry and The PULSE Institute, Stanford University, Stanford, California 94305, United States}}
\alsoaffiliation{SLAC National Accelerator Laboratory, Menlo Park, California 94024, United States}
\author{Lixin Lu}
\affiliation
{{Department of Chemistry and The PULSE Institute, Stanford University, Stanford, California 94305, United States}}
\alsoaffiliation{SLAC National Accelerator Laboratory, Menlo Park, California 94024, United States}
\author{Yuanheng Wang}
\affiliation
{{Department of Chemistry and The PULSE Institute, Stanford University, Stanford, California 94305, United States}}
\alsoaffiliation{SLAC National Accelerator Laboratory, Menlo Park, California 94024, United States}
\author{Todd J. Mart{\'i}nez}
\email{todd.martinez@stanford.edu; toddjmartinez@gmail.com}
\affiliation
{{Department of Chemistry and The PULSE Institute, Stanford University, Stanford, California 94305, United States}}
\alsoaffiliation{SLAC National Accelerator Laboratory, Menlo Park, California 94024, United States}

\begin{document}

	\maketitle

\begin{abstract}
Simulations of photochemical reaction dynamics have been a challenge to the theoretical chemistry community for some time. In an effort to determine the predictive character of current approaches, we predict the results of an upcoming ultrafast diffraction experiment on the photodynamics of cyclobutanone after excitation to the lowest lying Rydberg state (S$_2$). A picosecond of nonadiabatic dynamics is described with ab initio multiple spawning. We use both time dependent density functional theory and equation-of-motion coupled cluster for the underlying electronic structure theory. We find that the lifetime of the S$_2$ state is more than a picosecond (with both TDDFT and EOM-CCSD). The predicted UED spectrum exhibits numerous structural features, but weak time dependence over the course of the simulations. 
\end{abstract}
\newpage
\section{Introduction}
Accurate simulation of photorelaxation processes remains a challenge for modern computational chemistry. Photoexcitation of a chemical system is generally followed by significant structural reorganizations due to differences between the potential energy surfaces (PESs) of the excited and ground electronic states, often accompanied by radiationless transfer of population between electronic states. Such population transfer tends to occur around regions where PESs of different electronic states cross (so called `conical intersections'\citep{teller1937crossing,yarkony1996diabolical}) and stems from nonadiabatic processes due to the breakdown of the Born-Oppenheimer approximation. Photorelaxation pathways therefore tend to involve coupled electronic and nuclear dynamics that is challenging to model computationally. There have been many advances towards improved modeling of electronic excitations\citep{loos2018mountaineering,veril2021questdb,liang2022revisiting} and nonadiabatic molecular dynamics,\citep{curchod2018ab,crespo2018recent} but it can be difficult to quantify the accuracy that can be expected in reproducing experimental observables from time-resolved experiments. Of course, there are many successful studies\citep{timmers2019disentangling,wolf2019photochemical,yang2020simultaneous,zinchenko2021sub,ridente2023femtosecond} combining experiment and theory, and few would doubt the utility of theory and simulation in the interpretation of such experiments. Yet, the degree of predictiveness that can be expected from photochemical simulations remains largely unquantified.  Blind tests provide a means to begin to quantify the accuracy of photochemical simulations. 

\begin{figure}[htb!]
\centering
    \includegraphics[width=0.3\linewidth]{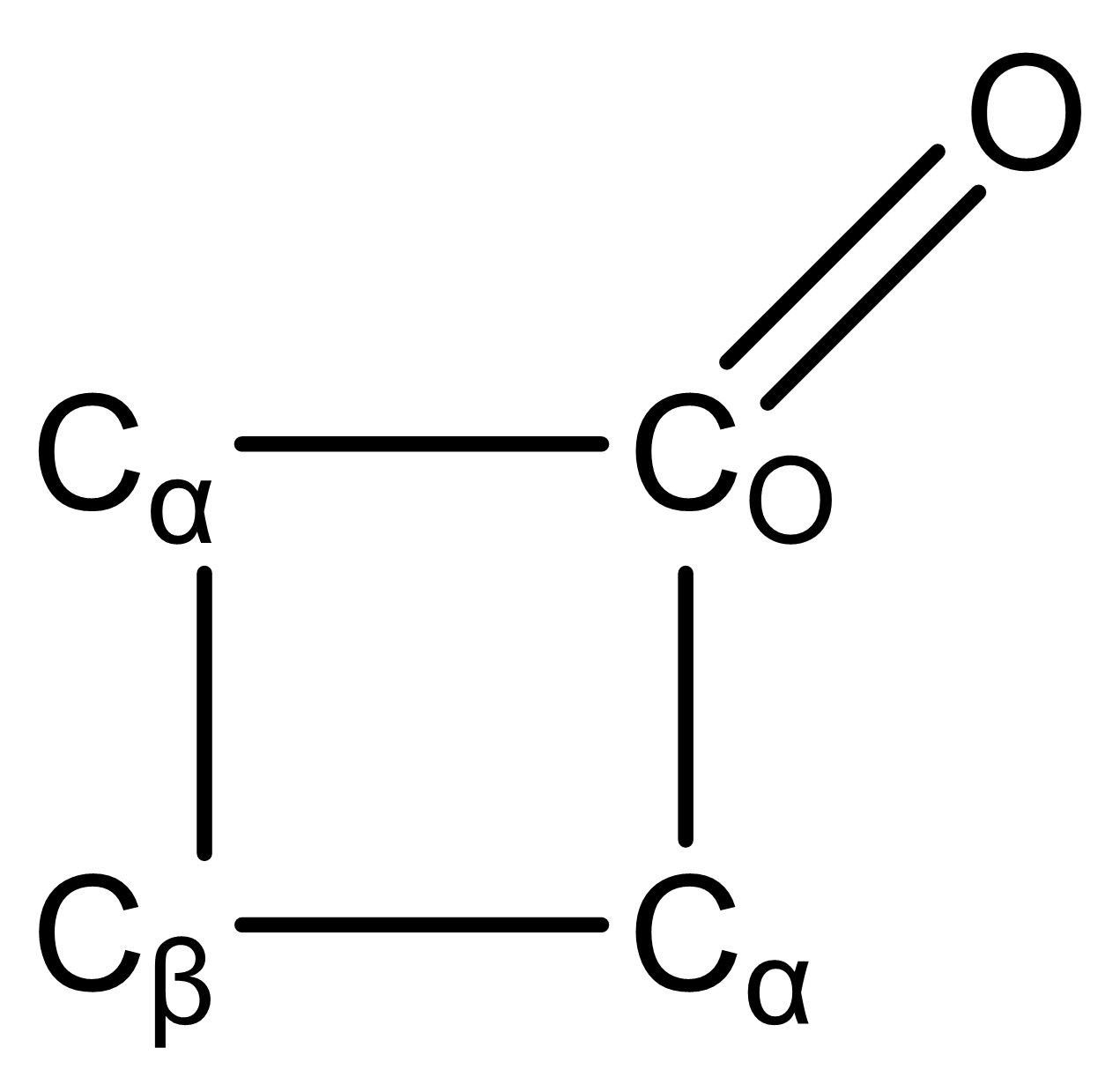}
\caption{Cyclobutanone, with different C atoms labeled (H atoms not shown for clarity).}
\label{fig:chemdraw}
\end{figure}

The photochemistry of cyclobutanone (Fig. \ref{fig:chemdraw}) was chosen as such a blind test for this special issue of The Journal of Chemical Physics. Photoexcitation of ketones often leads to Norrish reactions\citep{norrish1936photodecomposition} in which the bond between the carbonyl carbon (C$_O$) and an $\alpha$ carbon (C$_\alpha$) is homolytically cleaved. Cyclobutanone is additionally expected to have considerable ring strain (due to the presence of an unsaturated carbon in a four membered ring) and as such has the potential to undergo significant structural dynamics upon photoexcitation. Previous work\citep{diau2001femtochemistry,kao2020effects,xia2015excited,liu2016new} has investigated the photochemistry upon excitation to the first excited singlet (S$_1$) state, reporting ring opening and formation of products like carbon monoxide (CO), ethene (\ce{C2H4}) and ketene (\ce{H2C=C=O}). However, the S$_1$ state has a low absorption cross-section due to its n$\to\pi^*$ character at the ground (S$_0$) state equilibrium geometry. Therefore, exciting a substantial proportion of the molecules to the S$_1$ state is likely to require high pump power which can lead to multiphoton excitation/ionization. In order to avoid such phenomena, the experimental study associated with this blind test instead aims to use 200 nm light to excite cyclobutanone to the optically-bright second excited singlet (S$_2$) state with n$\to$3s Rydberg character. The excited molecules will be subsequently probed by ultrafast electron diffraction (UED\citep{centurion2022ultrafast}) to reveal the time-resolved structural dynamics occurring from photorelaxation. 

Certainly, the accuracy of the PESs is an important factor in describing excited state dynamics, and recently it has been suggested that this might often be critical.\citep{slavicek2023electronic} Unfortunately, the most reliable methods are generally too computationally intensive for first principles molecular dynamics on the necessary timescales. One way to circumvent this difficulty is to use computationally expensive methods (reliable with few or no adjustable parameters, but not easily applied within nonadiabatic dynamics) to calibrate and/or validate less expensive methods which can be used for long-time nonadiabatic dynamics simulations. In photochemical problems, this calibration and validation most naturally begins with the electronic absorption spectrum. The equation-of-motion coupled cluster (EOM-CC\cite{bartlett2007coupled}) method is often able to provide accurate results for molecules in their equilibrium geometries. The only ``parameters'' involved are the level of excitation (e.g. single and double excitations in EOM-CCSD\cite{stanton1993equation}) and the basis set. These parameters are reasonably well characterized, systematically improvable (for increased computational cost) and the level of convergence is relatively straightforward to test for, such as through the use of a higher level of coupled cluster theory like CC3\cite{koch1997cc3} or a larger basis set. When the experimental absorption spectrum is also available, direct comparison of the ``reference'' method with experiment allows for the assessment of any uncertainties. The absorption spectrum can then be computed with candidate methods (``lower level") for the dynamics to validate the accuracy of the latter. Of course, the ability to reproduce the absorption spectrum is only a necessary (and not sufficient) condition for the accuracy of the dynamics simulation. Validation of the dynamics carried out with the ``lower level'' method can be continued by identifying nearby critical points on the PES influencing the dynamics (such as excited state minima or minimal energy conical intersections) and comparing these (energetically and geometrically) to critical points optimized with the reference method. Ideally, this back-and-forth of computing dynamics and validating the potential energy surfaces continues throughout the dynamics, allowing one to bootstrap one's way to a reliable description of the photochemistry.

In some cases, time-dependent density functional theory (TDDFT\citep{casida1995time,dreuw2005single}) could be a reasonable choice for the lower level method, as it is computationally tractable for even large molecules. TDDFT and EOM-CCSD obtain excited states from the linear response of a ground-state DFT or CCSD solution to time-dependent electric fields, respectively. As a result, neither can describe the correct topology of conical intersections between the reference ground state and excited states.\citep{levine2006conical} However, they can describe conical intersections between excited states, as long as complications from any non-Hermitian eigenvalue problems (which can lead to complex energy eigenvalues) can be avoided. TDDFT in the Tamm-Dancoff approximation\citep{hirata1999time} (TDA-TDDFT) is guaranteed to avoid this problem by construction (although other problems can be encountered in regimes where the ground state reference starts to break down, such as highly stretched bonds\citep{hait2019beyond}). For EOM-CCSD, the problem has been studied in detail\citep{ hattig2005,kohn2007,kjonstad2017crossing} and formal solutions have been advanced.\citep{williams2023,kjonstad2017,kjonstad2019}

Since internal conversion out of the S$_2$ state of cyclobutanone is expected to involve conical intersections with the S$_1$ or S$_3$ states, TDDFT or EOM-CCSD should in principle be capable of describing the initial photorelaxation following 200 nm excitation (but would be highly questionable for modeling the S$_1$ $\rightarrow$ S$_0$ decay). The feature corresponding to the S$_2$ state in the experimental UV absorption spectrum of cyclobutanone \citep{keller2013mpi,udvarhazi1965vacuum} (centered at 194 nm) appears to exhibit vibrational fine structure, suggesting that there is a local minimum on S$_2$ with a significant lifetime. As a result, the initial photorelaxation of cyclobutanone following excitation with the 200 nm pump is likely to involve some localization of nuclear population about this S$_2$ minimum, potentially followed by internal conversion arising from vibrational motions approaching conical intersections. We thus start with these linear response methods to address the early time photodynamics of cyclobutanone. In this work, we will study the photorelaxation dynamics of cyclobutanone following photoexcitation by a 200 nm pulse using TDDFT and EOM-CCSD in conjunction with the \textit{ab initio} multiple spawning (AIMS) method\citep{ben2000ab,ben2002ab,curchod2018ab} for nonadiabatic molecular dynamics. Our results indicate that population transfer out of the S$_2$ state requires a few picoseconds under these conditions, leading to (relatively) slow structural dynamics. A complete prediction of the photochemistry would need to describe the decay from S$_1\to$ S$_0$ (which cannot be done with either EOM-CCSD or TDDFT), but we did not pursue this here because it seems unlikely that the signatures for this process will be visible in the experimental UED signal, due to the slow growth of S$_1$ population through internal conversion from S$_2$.

\section{Computational methods}\label{sec:comp_methods}
All (TD)DFT calculations utilized the LRC-$\omega$PBE functional\citep{rohrdanz2008simultaneous} (with the default value of $\omega=0.3$ bohr$^{-1}$) and the aug-cc-pVDZ basis set,\citep{dunning1989gaussian,kendall1992electron} being run with the GPU accelerated TeraChem software package.\citep{seritan2021terachem,ufimtsev2008quantum,ufimtsev2009quantum,ufimtsev2009quantumii} While TDDFT with local functionals generally systematically underestimates Rydberg state excitation energies,\citep{tozer1998improving} range separated hybrids like LRC-$\omega$PBE tend to fare much better for such states (having $\sim 0.3$ eV errors for Franck-Condon region vertical excitation energies\citep{liang2022revisiting}).
The chosen aug-cc-pVDZ basis was also sufficiently flexible to represent the Rydberg states, as indicated by prior work\citep{liang2022revisiting} and comparison to doubly augmented basis sets, as discussed in the supporting information (SI). TDA\citep{hirata1999time} was utilized to ensure that the linear-response eigenproblem is Hermitian and thereby guarantee real-valued excitation energies. 

EOM-CCSD calculations were run with the CPU based e$^{T}$ software package\citep{folkestad20201}, using the same aug-cc-pVDZ basis set. The frozen-core approximation was not utilized for the CC calculations. EOM-CCSD is known to be quite accurate for the valence and Rydberg excited states of small molecules, predicting vertical excitation energies at the Franck-Condon geometry to $\sim 0.1$ eV accuracy.\citep{loos2018mountaineering} It is nonetheless quite computationally demanding, with computational cost scaling formally as the sixth power of the system size.\citep{bartlett2007coupled} Although recent work has shown how to reduce this to quartic scaling in practice,\citep{hohenstein2012thc3,hohenstein2013thceom,hohenstein2019rrcc1,hohenstein2019rrcc2,hohenstein2022rrcc3} the analytic gradients and nonadiabatic couplings necessary for nonadiabatic dynamics simulations have yet to be implemented. The high cost of EOM-CC in conjunction with its known defective description of certain excited-state crossings\citep{ hattig2005,kohn2007,kjonstad2017crossing} has, to the best of our knowledge, prevented the use of EOM-CCSD based simulations of nonadiabatic dynamics with a proper treatment of conical intersections although it has been used to compute absorption spectra through the use of quasi-diabatization in the past\cite{baeck2003ab}. Some of the authors (O.J.F. and T.M.) have worked on developing an interface for performing AIMS simulations using EOM-CCSD with the e$^{T}$ package, enabling the first EOM-CCSD based nonadiabatic dynamics simulations that feature a proper description of conical intersections between excited states. A more detailed description of this interface, along with its promising application to the photochemistry of thymine is the subject of a forthcoming publication.\citep{kjonstad2024} 

All calculations use spin-restricted orbitals despite the potential for bond cleavage, as spin-contamination can lead to significant qualitative problems with the excited state PES predicted by linear-response methods.\citep{hait2019beyond}

\begin{figure}[htb!]
\begin{minipage}{0.48\textwidth}
    \centering
    \includegraphics[width=\linewidth]{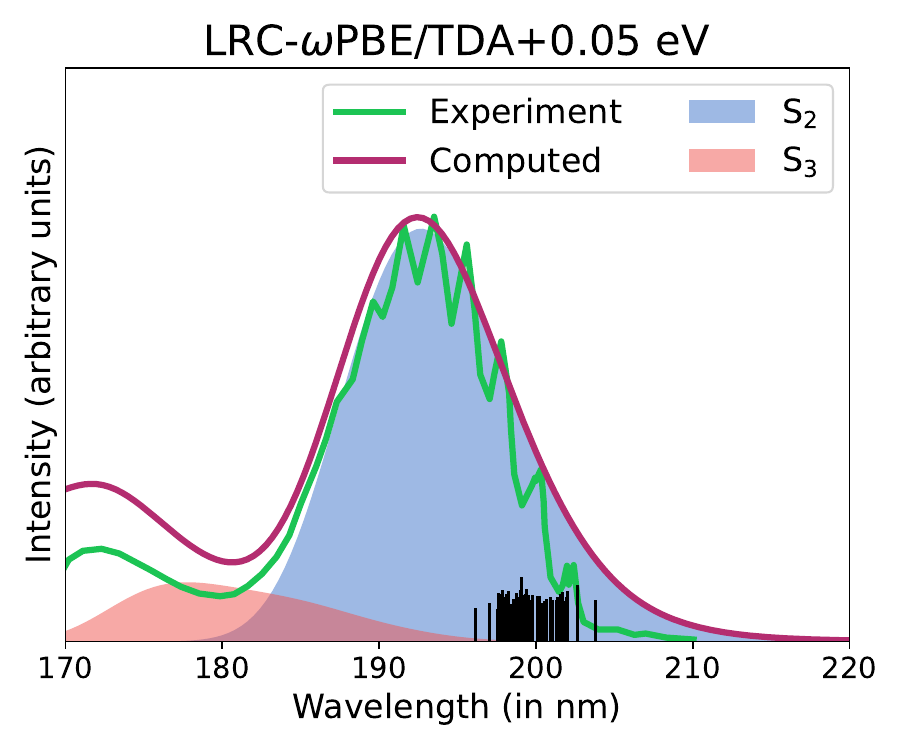}
\end{minipage}
\begin{minipage}{0.48\textwidth}
    \centering
    \includegraphics[width=\linewidth]{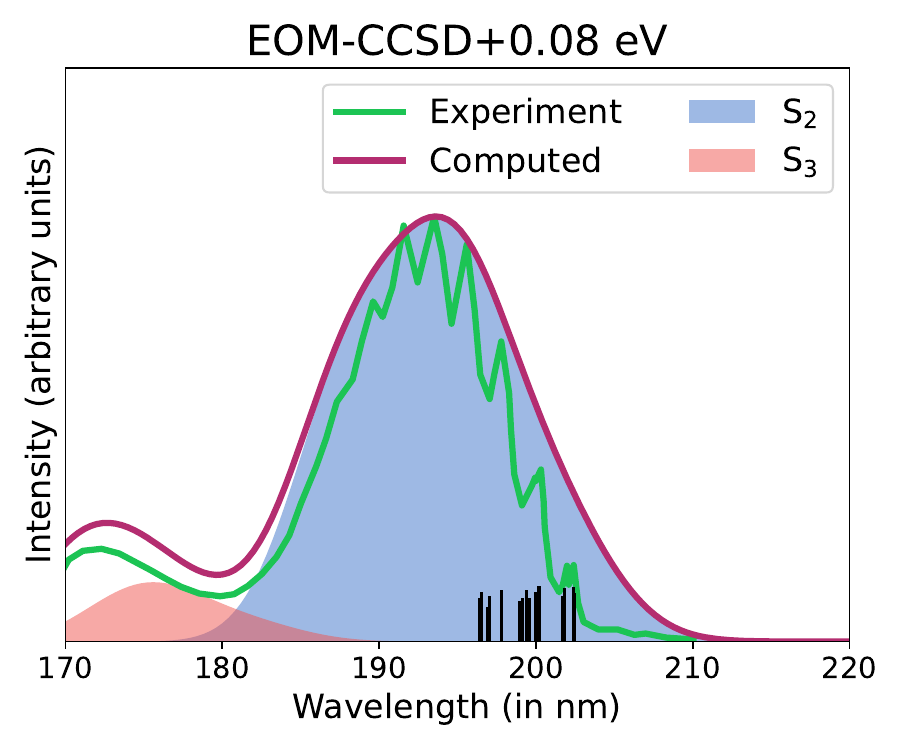}
\end{minipage}
\caption{Computed UV absorbtion spectrum from LRC-$\omega$PBE/aug-cc-pVDZ (left) and EOM-CCSD/aug-cc-pVDZ (right), compared to experiment. \citep{keller2013mpi,udvarhazi1965vacuum} The stick spectra for the sampled geometries (4096 for DFT and 100 for CCSD, sampled from the ground state harmonic vibrational wavefunction of the corresponding electronic structure method) were broadened by Gaussians with $\sigma=0.1$ eV and summed, followed by a constant energy shift (noted in the title) to align the maxima at $\sim 194$ nm with experiment. This semiclassical treatment is not expected to (and does not) reproduce the vibrational fine structure in the experimental spectrum. 
The contributions from the S$_2$ and S$_3$ states to the total computed spectrum is also indicated, showing that only the S$_2$ state has perceptible absorption at the 200 nm pump wavelength. The absorption energies for the geometries selected for the nonadiabatic simulations (see Computational Methods) are depicted as black lines with heights proportional to the corresponding oscillator strength.}
\label{fig:absorb}
\end{figure}

\subsection{Selection of Initial Conditions For AIMS Simulations}
Initial nuclear positions and momenta for the AIMS simulations were generated by sampling from the ground state harmonic vibrational wavefunction of the molecule about the S$_0$ state equilibrium geometry optimized with LRC-$\omega$PBE/aug-cc-pVDZ and CCSD/aug-cc-pVDZ for the TDDFT and EOM-CCSD dynamics, respectively. The UV absorbtion spectrum was constructed from the sampled geometries (see Fig. \ref{fig:absorb}), and aligned to the experimental spectrum\citep{keller2013mpi,udvarhazi1965vacuum} with a small constant energy domain blueshift (0.05 eV for TDDFT and 0.08 eV for EOM-CCSD, demonstrating the relative accuracy of the methods in question). This confirmed that the S$_2$ n$\to$3s Rydberg state was essentially the only state absorbing at the 200 nm pump wavelength and further validated both the TDDFT and EOM-CCSD methods in the Franck-Condon region. 

We subsequently obtained the initial nuclear positions for the AIMS simulations by sampling from the geometries utilized for the absorption spectrum calculations, weighted to approximate the effect of the 200 nm pump pulse. 
For this selection protocol, each geometry $i$ was assigned a weight $w_i=f_ie^{-\frac{(E_i-E_0)^2}{2\sigma^2}}$ where $f_i$ and $E_i$ are the oscillator strength and excitation energy (with the alignment shift included) of the S$_2$ state at this geometry, $E_0$ is the 200 nm pump energy (6.199 eV) and $\sigma_i=0.0517$ eV is a measure of the pump pulse width (corresponding to 4 nm full-width-at-half-max/FWHM). These weights $w_i$ were normalized to construct a discrete probability distribution for these geometries, from which a subset was selected (with replacement) for the AIMS dynamics simulations, to model the initial wavepacket after excitation with the 200 nm pump.  In total, 64 initial geometries (58 geometries sampled once, and 3 sampled twice, leading to 61 unique geometries) out of 4096 sampled for the absorption spectrum calculation were selected for the TDDFT/AIMS dynamics simulations, and 18 unique geometries for the EOM-CCSD/AIMS dynamics simulations. These geometries were paired with nuclear momenta sampled from the corresponding ground state harmonic vibrational wavefunction for the AIMS simulations. These initial geometry-momentum pairs are subsequently referred to as initial conditions for convenience.

\section{Results and Discussion}

\begin{figure}[htb!]
\begin{minipage}{0.49\textwidth}
    \centering
    \includegraphics[width=\linewidth]{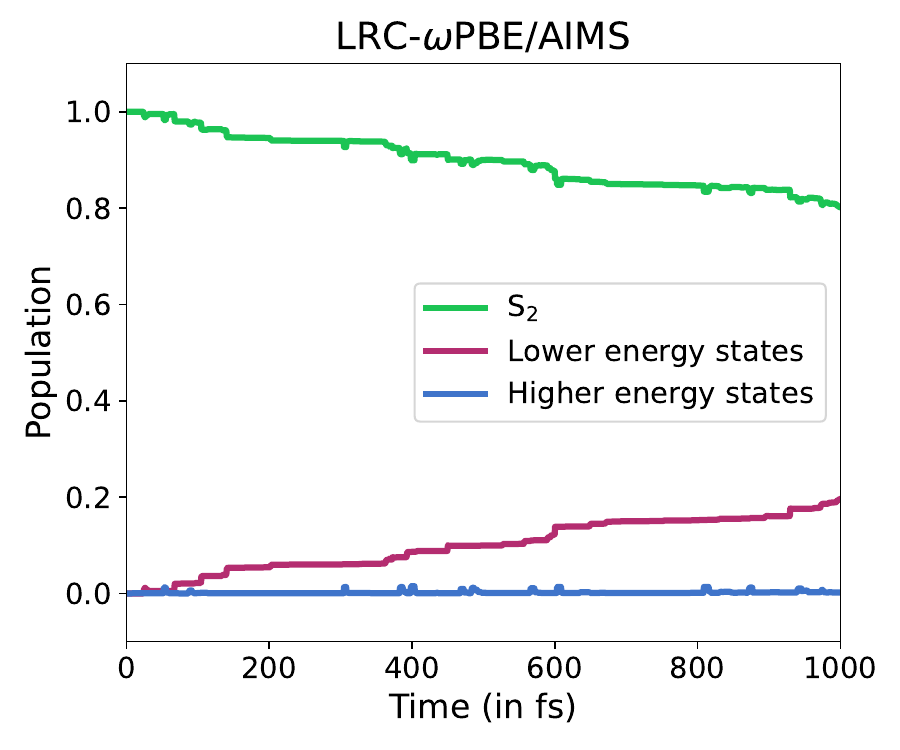}
\end{minipage}
\begin{minipage}{0.49\textwidth}
    \centering
    \includegraphics[width=\linewidth]{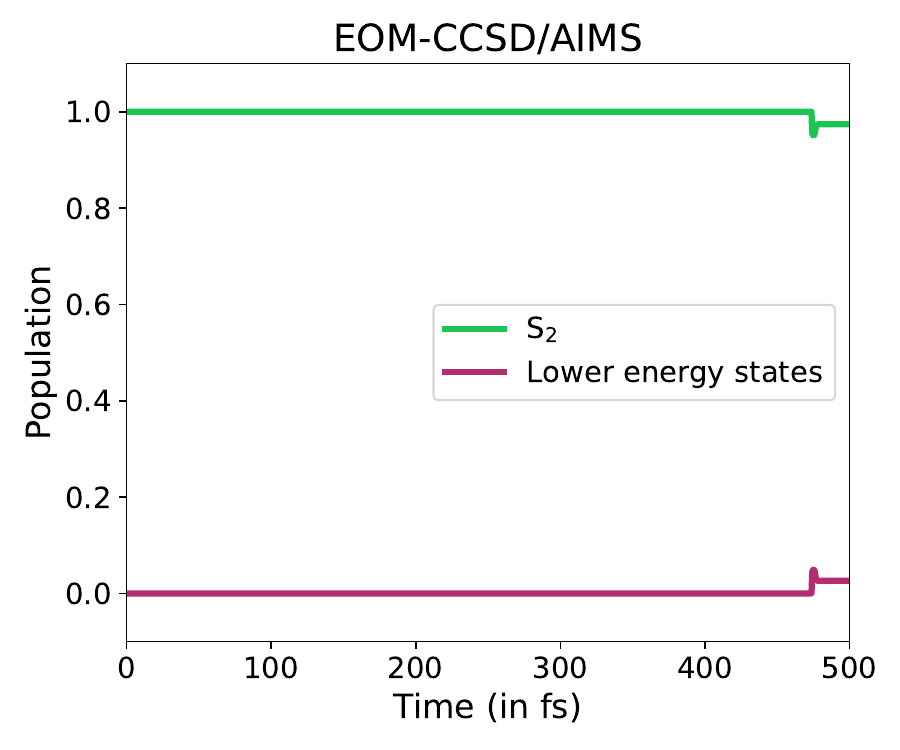}
\end{minipage}
\caption{Electronic state populations from AIMS simulations using LRC-$\omega$PBE (left) and EOM-CCSD (right). Note that the EOM-CCSD simulations were only run for 500 fs, while TDDFT results up to 1 ps are shown. One initial condition out of 64 was neglected for the TDDFT population plot as the dynamics could not be run to 1 ps (see SI for discussion). The non S$_2$ populations are described as belonging to lower energy states (i.e. S$_1$ and S$_0$) and higher energy states (i.e. S$_3$ and beyond) instead of precise state labels as the S$_1\to$ S$_0$ internal conversion cannot be modeled by the electronic structure methods utilized for the AIMS simulations.}
\label{fig:pops}
\end{figure}
\subsection{Population Transfer Dynamics}
The evolution of population in different electronic states over the course of the AIMS simulations is shown in Fig \ref{fig:pops}. One initial condition was neglected as the dynamics could not be run to a full picosecond (see SI for additional discussion). The TDDFT results with the remaining 63 initial conditions indicate relatively slow internal conversion out of S$_2$ to lower energy singlet states, and essentially no population transfer to higher energy states. We find that the S$_2$ population at 1 ps is $\sim 0.80 \pm 0.04$ from TDDFT/AIMS simulations over the 63 initial conditions utilized for Fig. \ref{fig:pops} (the uncertainty being the $1\sigma$ value estimated from bootstrap statistics\citep{efron1986bootstrap} over the initial conditions).  In particular, we note that only 18 of these 63 initial conditions exhibit \textit{any} S$_2\to$ S$_1$ internal conversion within 1 ps over the course of the AIMS simulations, transferring $68\%$ of the initial S$_2$ population associated with these 18 initial conditions ($20\%$ across all 63 initial conditions). If we further assume that the neglected initial condition would have undergone complete internal conversion out of S$_2$ by 1 ps (as seems reasonable, see discussion in SI), we would instead obtain an S$_2$ population range of $0.79\pm0.04$. Thus, TDDFT dynamics predict a lifetime for S$_2$ of at least 2 ps.  

The EOM-CCSD calculations are quite computationally demanding and therefore were only attempted for 18 initial conditions up to 500 fs. In this timespan, we observe only one S$_2\to$ S$_1$ spawning event (at 474 fs) that transfers $47\%$ of the S$_2$ population corresponding to that initial condition to S$_1$ ($2.6\%$ of the total  S$_2$ population over all 18 initial conditions). These results suggest even slower internal conversion than the TDDFT calculations.   
Additionally, 2 initial conditions out of 18 lead to elongated C$_\alpha-$C$_O$ bond lengths (in excess of 2.5 {\AA}) which cause ground state CCSD convergence failures (at 384 and 441 fs, respectively). Thus, we cannot exclude the possibility of population transfer from S$_{2}\to$ S$_{1}$ within 500 fs occurring in these two AIMS simulations.  We do not observe any other decay channels in the EOM-CCSD/AIMS dynamics, with the other 15/18 ICs remaining on S$_{2}$ for the full 500 fs. The limited number of initial conditions and the relatively short duration of the EOM-CCSD AIMS simulations make it extremely difficult to obtain precise estimates of the S$_2$ lifetime. Even complete population transfer out of S$_2$ for the two initial conditions with convergence failures within 500 fs would be consistent with a few-picosecond timescale for S$_2$ population loss (naively indicating at least a 3 ps lifetime within a monoexponential model), supporting the general conclusions from the TDDFT simulations. In short, both TDDFT and EOM-CCSD predict an S$_2$ lifetime of more than 2 ps, with the more accurate EOM-CCSD method predicting the longer lifetime of the two, albeit with much greater uncertainty. 

\begin{figure}[htb!]
    \includegraphics[width=0.7\linewidth]{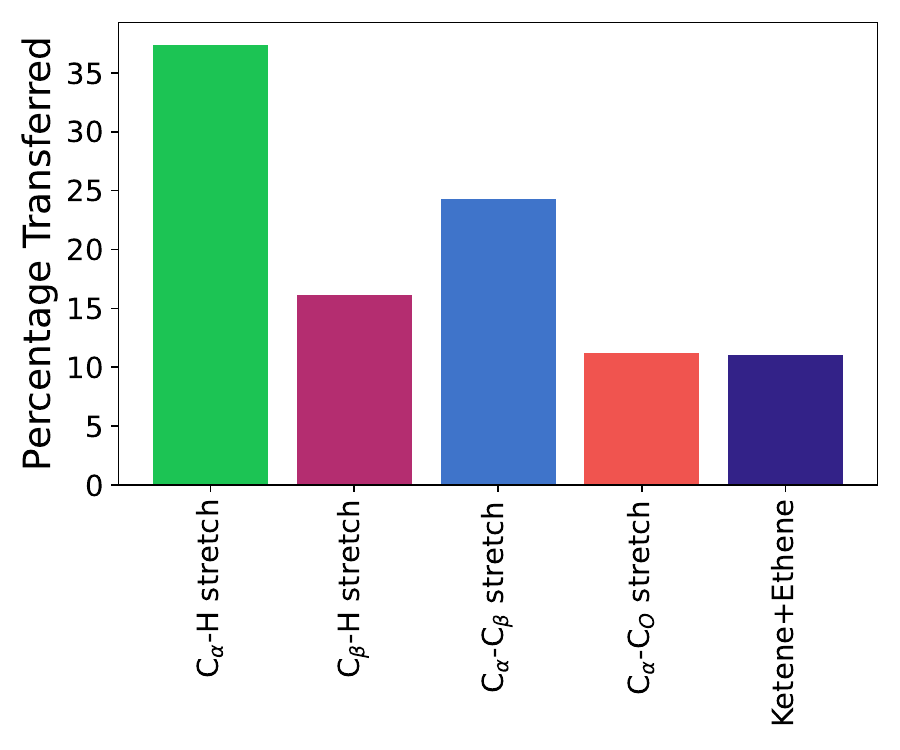}
\caption{Percentage of the total S$_2\to$ S$_1$ internal conversion within 1 ps in the LRC-$\omega$PBE/AIMS simulations, as classified via geometric distortions in the corresponding spawning geometries. We note that the total S$_2$ population transferred to S$_1$ is only 20\% over the 63 TDDFT/AIMS simulations, and the percentages shown in this figure correspond to fractions of this total transfer.}
\label{fig:IC_analysis}
\end{figure}

\subsection{Geometric Changes Associated With Internal Conversion}
The slow population transfer out of the S$_2$ state (in comparison to molecular vibrational timescales) indicates that most of the nuclear density is concentrated around minima on the S$_2$ potential energy surface, with internal conversion occurring as conical intersections with S$_1$ (and very rarely, S$_3$) are approached during vibrations away from the minima. We therefore attempt to determine the geometric distortions on the S$_2$ surface associated with spawning to S$_1$ within AIMS (representing regions of strong nonadiabatic coupling between the S$_2$ and S$_1$ states). Only 18 TDDFT initial conditions out of 63 lead to internal conversion to S$_1$ within 1 ps (transferring $\sim 20\%$ of the total S$_2$ population), and they appear to do so through a limited set of geometric distortions at the spawning geometry, as shown in Fig \ref{fig:IC_analysis}. Stretching of a C$_\alpha-$H bond is responsible for the largest fraction of internal conversion, representing 37\% of the S$_2$ population that is transferred to S$_1$. Stretching of C$_\beta-$H bonds on the other hand, appears to only contribute 16\% of the internal conversion. These pathways with relatively little distortion of the four membered ring thus contribute to about half of the internal conversion, and appear to involve S$_2$/S$_1$ states corresponding to excitations out of the O lone pair to states of mixed valence/Rydberg character. 

The remainder of internal conversion ($\sim 46\%$) occurs through states with significant C-C bond elongation in the ring, such as stretching of a C$_\alpha-$C$_\beta$ bond (24\%), stretching of a C$_\alpha-$C$_O$ bond (11\%) or both, leading to structures resembling [2+2] cycloelimination of ketene and ethene (11\%).  These tend to involve S$_2$/S$_1$ states of valence character, typically with excitations out of the stretched C-C formally $\sigma$ bonding orbitals as well as the O lone pair into antibonding levels. 

Interestingly, the only spawning event to occur in the EOM-CCSD/AIMS simulations involves the stretching of a C$_\alpha-$C$_O$ bond on the S$_2$ surface, which is the distortion also observed for the two cases where the simulations fail due to ground state CCSD convergence failures. In contrast, TDDFT/AIMS shows only a single spawning event involving C$_\alpha-$C$_O$ stretch within 500 fs, transferring $6\%$ of the total population converting to S$_1$ within this time. 
The relative propensity for C$_\alpha-$C$_O$ bond elongation in the EOM-CCSD calculations (relative to other structural distortions that could permit internal conversion) thus appears to be higher than in TDDFT. 

\begin{figure}[htb!]
    \includegraphics[width=\linewidth]{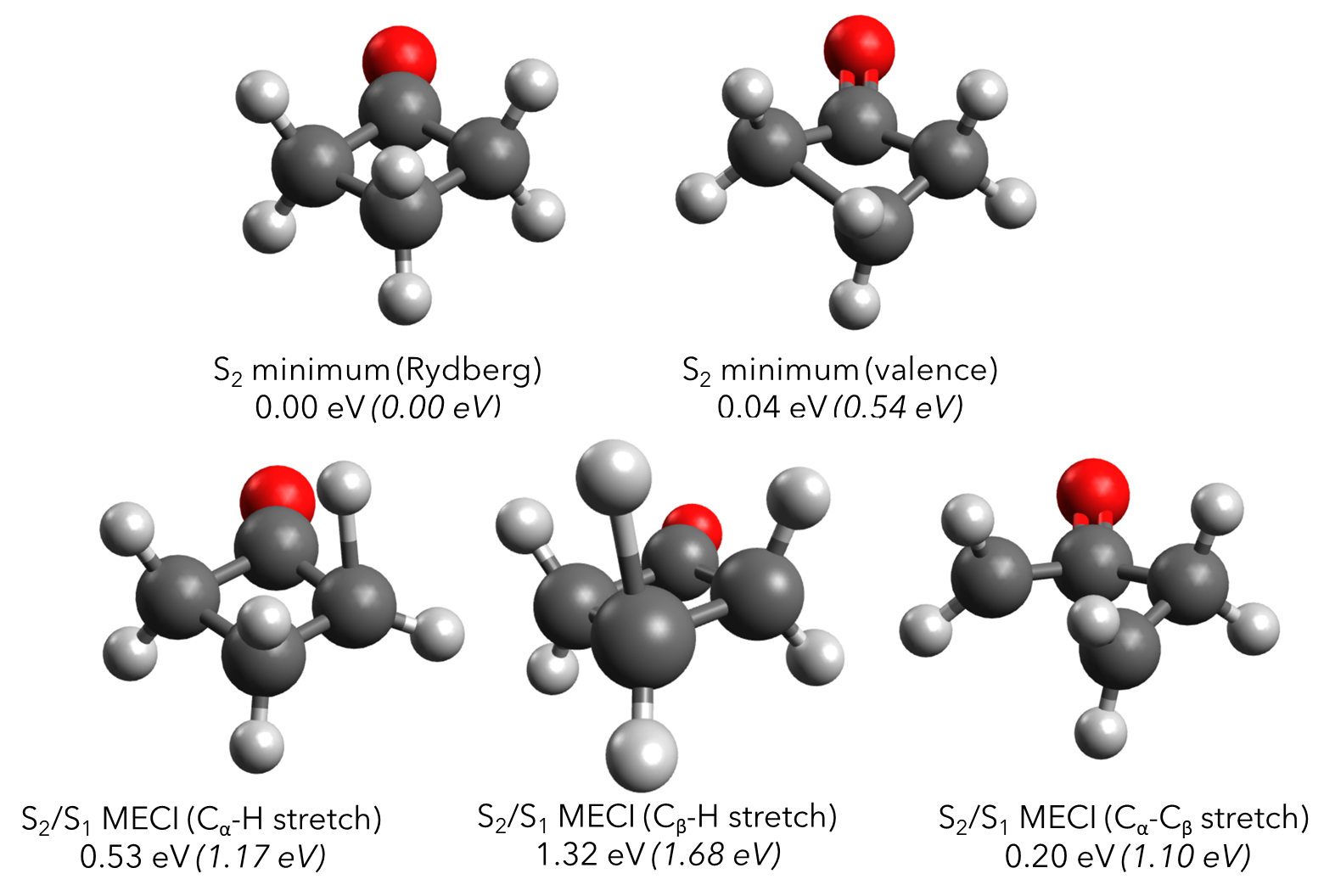}
\caption{Energy minima on the S$_2$ surface and minimum energy conical intersections (MECIs) between S$_2$ and S$_1$ as optimized by TDDFT. Relative energies from TDDFT (as well as from EOM-CCSD on the corresponding structures optimized with EOM-CCSD, given inside parentheses in italics) are also provided.}
\label{fig:structs}
\end{figure}

\subsection{Critical Points on Potential Energy Surfaces}
While the ultrafast dynamics of photorelaxation in chemical systems are typically nonequilibrium, general features about the mechanisms can be classified or interpreted on the basis of critical points like local minima or minimum energy conical intersections (MECIs). Both TDDFT and EOM-CCSD find two separate types of local minima on the S$_2$ surface, as shown in Fig \ref{fig:structs}. The lower energy structure is of Rydberg character and involves relatively little distortion relative to the Franck-Condon geometry, suggesting this form is responsible for the vibrational fine structure in the UV absorption spectrum. The higher energy structure involves an elongated C$_\alpha-$C$_\beta$ bond, and is of valence excitation character (out of the stretched bond $\sigma$ orbital to the $\pi^*$ level). These two minima are energetically separated by only 0.04 eV at the TDDFT level, while EOM-CCSD predicts that the valence minimum is higher in energy by 0.54 eV relative to the Rydberg minimum. Interestingly, both TDDFT and EOM-CCSD appear to predict low energy configurations with bond dissociation (through either cleavage of a C$_O-$C$_\alpha$ bond or cycloelimination to form ketene+ethene) that are below the aforementioned local minima on the S$_2$ surface, but precise minimum energy structures corresponding to these geometric changes could not be optimized as the single reference TDDFT/EOM-CCSD calculations become unreliable for highly stretched bonds. The low rate of C$_O-$C$_\alpha$ bond breaking in the TDDFT/AIMS simulations nonetheless suggest that nonnegligible effective barriers likely exist towards the formation of these low energy bond cleaved structures at the TDDFT level. In this regard, it is worth noting that the energy released by relaxation on the S$_2$ surface from the Franck-Condon point to Rydberg character minimum is 0.24 eV with TDDFT and 0.30 eV with EOM-CCSD. As the 200 nm pump pulse is lower in energy than the experimental absorption band maximum ($\sim$ 194 nm) by about $\sim 0.2$ eV, even less energy is made available by geometrical relaxation on the S$_2$ surface to the Rydberg character minimum. However, the semiclassical modeling of nuclear dynamics via 
initial conditions sampled from the harmonic ground state wavefunction provides the system with additional energy (from zero-point energy) that could help drive processes that would appear to be forbidden from relative energetics of critical point geometries alone.     

We also find three distinct MECIs between the S$_2$ and S$_1$ surfaces with TDDFT (shown in Fig \ref{fig:structs}), roughly corresponding to the two distinct S$_2$ minima (as shown in Fig \ref{fig:structs}). The two MECIs that involve elongation of C-H bonds geometrically resemble the Rydberg character S$_2$ minimum, while the third MECI resembles the valence S$_2$ minimum but with further stretching of the C$_\alpha-$C$_\beta$ bond. These MECIs could not be directly found at the EOM-CCSD level as the method involves diagonalization of a non-Hermitian matrix that led to complex eigenvalues in the neighborhood of these conical intersections.\citep{kjonstad2017crossing} The minimum energy geometries obtained from a CI search with $\le 0.02$ eV S$_2$/S$_1$ energy gaps were consequently treated as ``approximate MECIs" for estimating relative energies at the EOM-CCSD level (also reported in Fig \ref{fig:structs}).   

The TDDFT/AIMS simulations indicate that the MECI involving C$_\alpha$-H stretching is responsible for roughly 37\% of the population transfer out of S$_2$ within 1 ps (see Fig. \ref{fig:IC_analysis}). The MECI with C$_\beta$-H elongation is substantially higher in energy (1.32 eV vs the Rydberg character minimum at the TDDFT level) and as such is perceptibly less accessible, accounting for 16\% of the S$_2\to$ S$_1$ internal conversion in the TDDFT/AIMS simulations. The EOM-CCSD ``approximate MECIs" between S$_2$ and S$_1$ surfaces with similar geometries  appear to be even less energetically accessible than what is indicated by TDDFT, consistent with the lack of internal conversion through C-H elongation channels in the EOM-CCSD/AIMS simulations. 

The valence character MECI involving stretching of the C$_\alpha$-C$_\beta$ bond is the lowest of the three MECIs in energy (0.20 eV vs the Rydberg minimum at the TDDFT level), but involves a change in S$_2$ excitation character from Rydberg to valence, suggesting the presence of an intermediate barrier. On the other hand, a nudged elastic band (NEB\citep{jonsson1998nudged}) calculation indicated that no barrier exists between the valence minimum and this MECI, the path being purely uphill in energy. In this regard, we note that there is a S$_2$/S$_3$ MECI that can be found along the NEB path connecting the two S$_2$ minima at the TDDFT level, which is 0.59 eV above the Rydberg minimum at the TDDFT level. A proper (i.e. non-defective) MECI between S$_2$ and S$_3$ can similarly be obtained at the EOM-CCSD level, being 0.93 eV above the Rydberg S$_2$ minimum.  The relaxation through the stretching of the C$_\alpha$-C$_\beta$ bond starting from the Rydberg chracter S$_2$ minimum can therefore be seen as a two step process involving traversing the S$_2\to$ S$_3$ MECI to access the valence character minimum, followed by an uphill climb to this S$_2\to$ S$_1$ MECI. 
Nonetheless, higher energy pathways involving adiabatic change of S$_2$ excitation character between the two S$_2$ minima potentially also exist. As a result of these effective barriers, this MECI is responsible for a relatively small proportion of the S$_2\to$ S$_1$ internal conversion at the TDDFT/AIMS level (24\%) and appears to not be accessed at the EOM-CCSD/AIMS simulations. 

Fig. \ref{fig:IC_analysis} shows that $\sim$ 22\% of the internal conversion from S$_2\to$ S$_1$ within 1 ps at the TDDFT level occurs via structures with elongated C$_O-$C$_\alpha$ bonds or cycloelimination to form ketene+ethene. 
We could not optimize S$_2$ minima structures or S$_2$/S$_1$ MECIs corresponding to these channels with either TDDFT or EOM-CCSD due to difficulties associated with converging the reference ground state calculations in the stretched bond regime, but the S$_2$ energies that arose over the course of such optimizations were often lower than the local minima mentioned earlier. As a result, effective barriers must exist between the Rydberg S$_2$ minima where nuclear density accumulates per the AIMS simulations, and low energy stretched bond regimes on the S$_2$ PES. We did find a transition state structure corresponding to C$_O-$C$_\alpha$ bond elongation relative to the Rydberg character local minimum on the S$_2$ surface with TDDFT, that was 0.53 eV above that minimum. Interestingly, the ground state DFT solution at this transition state geometry is stable against spin-polarization, suggesting that TDDFT is qualitatively adequate for this degree of  C$_O-$C$_\alpha$ bond elongation (2.13 {\AA}). Constrained optimizations with EOM-CCSD indicated the presence of a barrier of $\sim 0.27$ eV against the elongation C$_O-$C$_\alpha$ (corresponding to a structure with a 2.075 {\AA} C$_O-$C$_\alpha$ stretch). This lower barrier from EOM-CCSD is consistent with the higher propensity to encounter C$_O-$C$_\alpha$ stretched geometries from the EOM-CCSD/AIMS simulations compared to TDDFT/AIMS, as discussed earlier. The relaxation pathway involving such stretched bond configurations is likely to entail rapid S$_1\to$ S$_0$ interconversion due to reduced excitation energies and ground state biradical character in the stretched bond regime, but this cannot be modeled with either TDDFT or EOM-CCSD. Nonetheless, a general photorelaxation mechanism involving C$_O-$C$_\alpha$ bond elongation from the Rydberg S$_2$ minimum leading to internal conversion and significant structural dynamics appears plausible for 200 nm excited cyclobutanone under experimental conditions. Both TDDFT/AIMS and EOM-CCSD/AIMS simulations appear to suggest a picosecond scale relaxation along C$_O-$C$_\alpha$ elongation based pathways, with TDDFT/AIMS indicating that other internal conversion pathways are more likely. However, the associated single reference electronic structure methods may not be sufficiently accurate for highly stretched C$_O-$C$_\alpha$ geometries, and the true rate may be either faster or slower depending on the true barrier along this channel.  

\subsection{Mechanistic Illustration}

\begin{figure}[htb!]
    \begin{minipage}{0.8\textwidth}
    \centering
        \includegraphics[width=\linewidth]{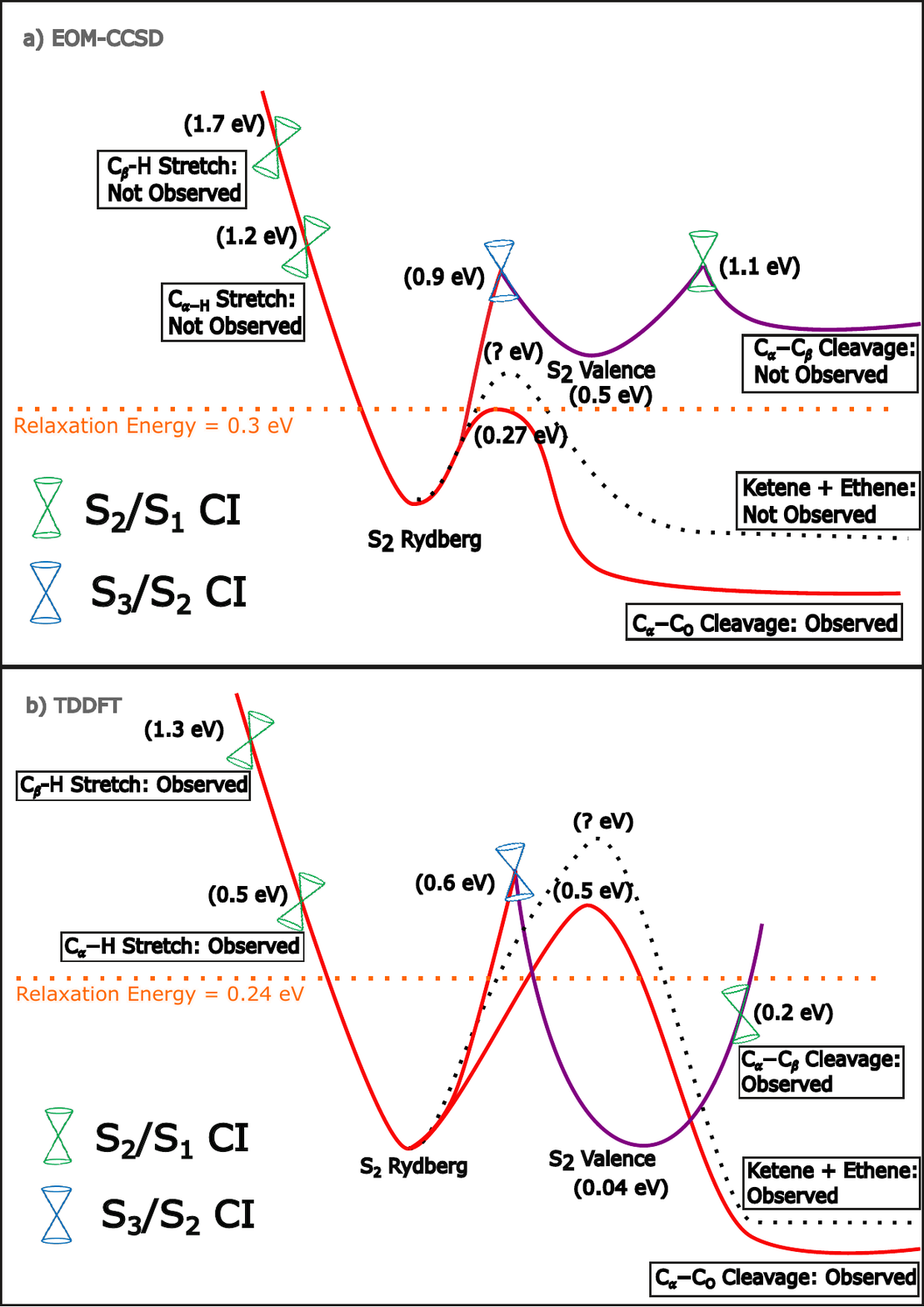}
    \caption{Schematic illustration of the mechanistic pathways observed (and possible, but not observed) in the EOM-CCSD/AIMS (a) and TDDFT/AIMS (b) dynamics. Note that the vertical, potential energy axis is approximate and not necessarily to scale. }
    \label{fig:mechanism_illustration}
    \end{minipage}
\end{figure}

We summarize the photodecay channels observed in our simulations in Fig. \ref{fig:mechanism_illustration}. Upon vertical excitation from the Franck-Condon geometry, we have a relaxation energy of 0.30 eV in EOM-CCSD and and 0.24 eV in TDDFT, reflecting the difference in S$_2$ energy at the Franck-Condon and S$_2$ minimum (Rydberg) geometries. For EOM-CCSD, this initial excess energy, while very close to the $\sim 0.27$ eV barrier to C$_O$-C$_\alpha$ elongation from the Rydberg-type minimum on S$_2$, is much less than the four MECIs (three S$_2\to$ S$_1$ shown in Fig. \ref{fig:structs} and one S$_2\to$ S$_3$) implicated in the TDDFT dynamics, with the lowest-lying of these being an S$_3$/S$_2$ MECI connecting the two S$_2$ minima. The C$_O$-C$_\alpha$ elongation pathway has by far the most readily surmountable barrier at the EOM-CCSD level and this is manifested in our dynamics, where we only observe distortions away from the S$_2$ Rydberg-type minimum in the form of C$_O$-C$_\alpha$ elongation (leading to one spawning event and two ground state convergence failures). Because we have only very limited dynamics at the EOM-CCSD/AIMS level, we cannot make any definitive statements about the channels that are not observed. Based on EOM-CCSD single point calculations at the various critical geometries, and their lack of participation in the EOM-CCSD/AIMS dynamics, we have sketched the higher lying part of the S$_2$ potential surface in Fig. \ref{fig:mechanism_illustration}a. We stress that this is a schematic diagram, intended to summarize the observed dynamics and critical points that we have located.

On the other hand, there is no pathway immediately accessible from the S$_2$ Rydberg-type minimum at the TDDFT level, with both the C$_\alpha$-H stretch S$_2$/S$_1$ and S$_3$/S$_2$ MECIs lying around 0.5-0.6 eV about the S$_2$ Rydberg-type minimum. In fact, the initial relaxation energy of 0.24 eV is insufficient to overcome any of the barriers leading away from the Rydberg-type minimum. However, the initial nuclear kinetic energy cannot be ignored (see Fig. S1) and this explains the observed population transfer events from S$_2\to$ S$_1$ in the vicinity of each type of MECI outlined in Fig. \ref{fig:IC_analysis}, with C$_\alpha$-H elongation being the predominant channel ($\sim 37\%$), C$_\alpha$-C$_\beta$ elongation next ($\sim 25 \%$), and C$_\beta$-H elongation, C$_\alpha$-C$_O$ elongation, and ketene and ethene being less prominent, each around $\sim 10-15 \%$. In the mechanistic cartoon for the TDDFT/AIMS simulations, we see that the C$_\alpha$-H MECI is accessible directly from the S$_2$ Rydberg minimum, while the C$_\alpha$-C$_\beta$ cleavage pathway, the next most observed, requires first traversing the barrier separating the two S$_2$ minima, before again moving slightly uphill away from the S$_2$ valence-type mimimum toward the C$_\alpha$-C$_\beta$ elongation MECI. Hence, even though the barriers leading away from the S$_2$ Rydberg-type mimimum are similar for these two pathways, the C$_\alpha$-C$_\beta$ elongation pathway requires a larger distortion and therefore is observed in lesser proportion in the TDDFT dynamics. The three remaining pathways (C$_\beta$-H elongation, C$_\alpha$-C$_O$ elongation, ketene and ethene) are the smallest contributors to the TDDFT dynamics, and are accordingly sketched as having the highest barriers. However, due to their low occurrence, a definitive ordering of the barriers is not attempted and the one suggested in Fig. \ref{fig:mechanism_illustration} should not be construed as such; rather, it is based on the single point TDDFT calculations at the critical geometries listed earlier to show the variety of channels that may be implicated in the photorelaxation dynamics of cyclobutanone upon excitation to S$_2$.

\begin{figure}[htb!]
    \includegraphics[width=\linewidth]{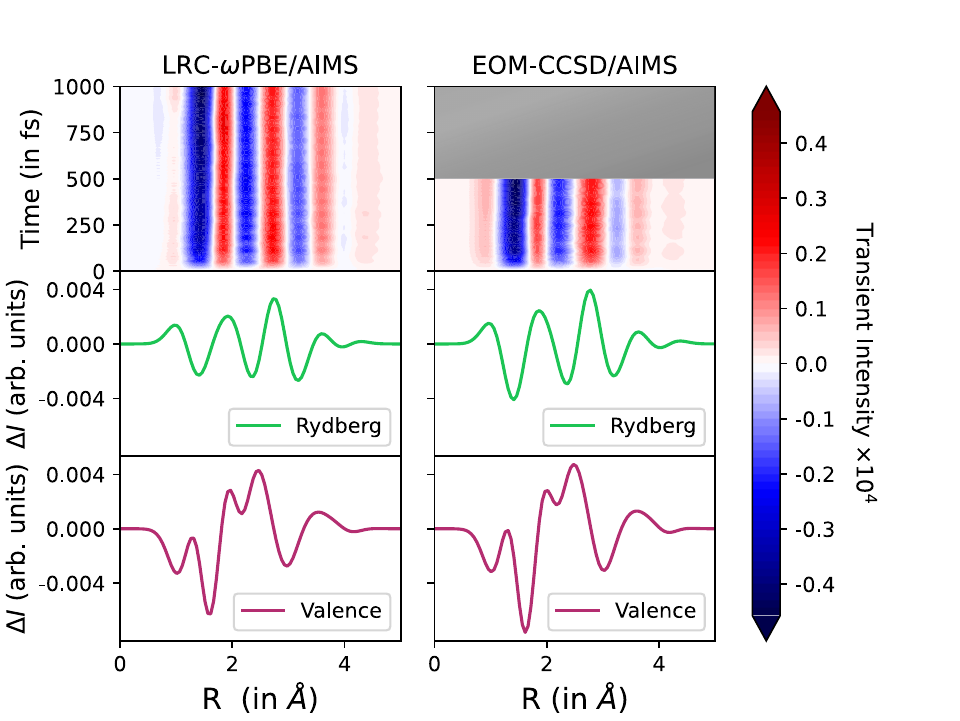}
\caption{Top row: Real-space UED transient spectra generated using LRC-$\omega$PBE/AIMS (left, 1 ps) and EOM-CCSD/AIMS (right, 500 fs). The transient signal was convolved in the time and spatial domains with Gaussians of FWHM 150 fs and 0.5 {\AA} respectively,  to better represent experimental conditions. Middle row: The difference in UED signals between the S$_2$ local minimum geometry of Rydberg character and the Franck-Condon geometry from LRC-$\omega$PBE (left) and EOM-CCSD (right). Bottom row: The difference in UED signals between the S$_2$ local minimum geometry of valence character and the Franck-Condon geometry from LRC-$\omega$PBE (left) and EOM-CCSD (right). }
\label{fig:UED}
\end{figure}

\subsection{Ultrafast Electron Diffraction}

Real-space transient ultrafast electron diffraction (UED) spectra arising from the AIMS simulations are shown in Fig. \ref{fig:UED} (momentum space diffraction patterns can be found in the SI). The UED spectra were obtained using an independent-atom model (IAM) and atomic UED cross sections generated by the ELSEPA program\citep{salvat2005elsepa} for a 3.7 MeV e$^-$ beam. The IAM utilizes only geometric information (and not electronic state charge densities), and cannot distinguish between different electronic characters for a given molecular structure. The computed transient UED signal was convolved with a Gaussian of 150 fs full-width half max (FWHM) in time to reflect the resolution of the instrument and a Gaussian of 0.5 {\AA} FWHM in space.

The main depletion (blue) and growth (red) features in the time-resolved UED spectrum from the AIMS simulations (top row of Fig. \ref{fig:UED}) appear to correspond very well with the difference signal (middle row of Fig. \ref{fig:UED}) computed from the Rydberg character S$_2$ minimum and the Franck-Condon geometry at the corresponding level of electronic structure. Furthermore, the positions of the transient features appear remarkably stable over time and no new significant features appear or disappear. This indicates a lack of significant structural dynamics during the course of the simulations. Lineouts taken at real-space separations corresponding to local maxima (minima) of the difference signal show a rapid rise (fall) from no net signal at $t=0$, followed by oscillations about the asymptotic long-time value (as shown in the SI). Fitting these lineouts to the sum of an exponential and a sinusoid indicate that the initial rapid rise/fall occurs in about $\sim 40-80$ fs, and the oscillation period is $\sim 65$ fs from both TDDFT/AIMS and EOM-CCSD/AIMS. The first timescale likely reflects relaxation to the Rydberg character local minimum following photoexcitation to S$_2$, and the second is consistent with ring stretching modes around this excited state minimum ($\sim$ 500 cm$^{-1}$; see Table S4). We note that the oscillations persist over our AIMS simulation timescales, and therefore might be experimentally observable with sufficient time resolution. In passing, we note that the difference signal between the valence character S$_2$ minimum and the Franck-Condon geometry aligns perceptibly less well with features in the transient signal, indicating lack of any significant buildup of nuclear density around such minima on the S$_2$ PES. 

It is interesting to consider how the experimental UED signal would differ if the S$_2\to$ S$_1$ internal conversion dynamics was faster or slower than the few ps timescale indicated by our AIMS simulations. If the S$_2$ lifetimes were large relative to the timescales of the UED experiment, the same stable set of transient features shown in the top panels of Fig. \ref{fig:UED} would be expected to persist to longer times. Vibrational coherence may be lost over time through intramolecular vibrational redistribution leading to loss of explicit oscillations in the transient UED signal and some additional broadening, but qualitatively the behavior should largely be expected to be similar to what is observed in Fig. \ref{fig:UED}, namely the nuclear density remaining localized about the Rydberg character S$_2$ minimum. 

Conversely, a faster timescale of internal conversion that leads to dissociative dynamics is likely to lead to a disappearance of some transient features observed in Fig. \ref{fig:UED}. In particular, it seems reasonable to expect that the short distance growth features ($R\le 2 $ {\AA}) might disappear over time as cyclobutanone dissociates into smaller products. A rate for internal conversion out of S$_2$ leading to larger structural dynamics thus might be experimentally discernable from the decay rate of the lineout corresponding to the growth feature that is predicted to be $R\sim 1.8$ {\AA} in the transient signal, as it is the lowest $R$ prominent growth feature observed in the computed transient UED spectrum.  

\subsection{Time Resolved Photoelectron Spectra}

In order to continue validating our simulation results, it would be prudent to compare to any closely related experiments. In fact, a time-resolved photoelectron spectroscopy\citep{stolow2004femtosecond} (TRPES) experiment has been previously reported for cyclobutanone after 200 nm excitation (using a 352 nm probe pulse, i.e. 3.5 eV).{\citep{kuhlman2012trpes} TRPES is a powerful technique for studying nonadiabatic dynamics and previous work has shown how to model TRPES spectra with AIMS.\citep{hudock2007trpes,glover2018butadienetrpes} We have therefore computed the TRPES signal arising from our AIMS simulations at the 352 nm probe wavelength, considering only one-photon ionization channels and following the previously described procedure.\citep{hudock2007trpes,glover2018butadienetrpes} Only the TDDFT/AIMS simulations were utilized for modeling TRPES, due to the limited number of initial conditions used for the EOM-CCSD/AIMS simulations.

Single photon ionization from the S$_0$ state of cyclobutanone is energetically forbidden at the 352 nm probe wavelength, and TRPES signal can only arise from electronic excited states. Furthermore, only the electronic ground state of the cation (the lowest doublet or D$_0$ state) can be energetically accessed with this probe. Our calculations therefore only considered ionization to the D$_0$ state from singlet excited states of neutral cyclobutanone. 

The kinetic energy of the ionized electron was computed as the difference between the energy of a 352 nm probe photon and the time-dependent ionization potential of the photoexcited molecule (the D$_0$ state cation energies being obtained from unrestricted LRC-$\omega$PBE/aug-cc-pVDZ calculations). 
Computed ionization potentials were shifted by -0.11 eV to correct for errors in the TDDFT vertical S$_0$/S$_2$ excitation and S$_0$/D$_0$ ionization energies. The probability of ionization is approximated by the S$_n$/D$_0$ Dyson orbital norm\citep{oana2007dyson} (where S$_n$ is the electronic state corresponding to a given trajectory basis function) when the computed electron kinetic energy is positive, and otherwise vanishes. We computed Dyson orbital norms at a number of geometries from the dynamics and found that these exhibited little variation. Thus, we approximate these as constant.
For each AIMS initial condition, ionization potentials were computed every 40 fs. To match the experimental cross-correlation between pump and probe, a Gaussian convolution of 124 fs FWHM in time was applied to the TRPES signal. An energy-domain Gaussian convolution of 0.02 eV FWHM was also applied to smooth the spectrum.

\begin{figure}[htb!]
\begin{minipage}{0.60\textwidth}
    \centering
    \includegraphics[width=\linewidth]{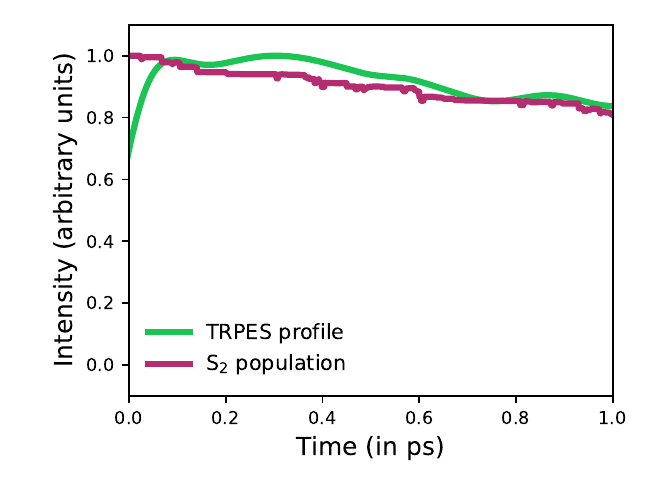}
\end{minipage}
\caption{Computed energy-integrated TRPES profile obtained from the first 1.0 ps of TDDFT/AIMS simulations, compared to the S$_2$ state electronic population from the same simulations.}
\label{fig:TRPES_TDDFT_1ps_allspawn}
\end{figure}

Fig.~\ref{fig:TRPES_TDDFT_1ps_allspawn} shows the energy-integrated TRPES profile obtained from the TDDFT/AIMS simulations, and compares it with the corresponding S$_2$ state population. The decay of the computed total integrated TRPES signal intensity tracks quite well with the S$_2$ electronic population from the underlying TDDFT/AIMS simulations. This indicates that the TRPES signal decay rate is a faithful measure of the internal conversion timescale, and is not significantly affected by ionization potential shifts during relaxation on S$_2$ (a possibility which is known to complicate the interpretation of TRPES experiments\citep{tao2011ethylene}).

The experimental study\cite{kuhlman2012trpes} reported that the total energy integrated TRPES signal was fit with a biexponential decay process with time constants of 0.31$\pm$0.06 ps and 0.74$\pm$0.02 ps. This is appreciably faster than the signal decay rate observed from our computational results in Fig.~\ref{fig:TRPES_TDDFT_1ps_allspawn}, where 80\% of the integrated signal persists at 1 ps. This disagreement and the demonstration in Fig. \ref{fig:TRPES_TDDFT_1ps_allspawn} that the TRPES signal is a faithful proxy for the S$_2$ population suggests that the S$_2$ lifetime from our TDDFT/AIMS simulations is too long. Of course, the EOM-CCSD/AIMS simulations predict an even longer S$_2$ lifetime, albeit with far more limited statistics. We can already anticipate that the measured UED signal may reflect this discrepancy.

\section{Conclusion}
In conclusion, our AIMS simulations employing TDDFT and EOM-CCSD methods indicate that internal conversion out of the S$_2$ state of cyclobutanone following photoexcitation by a 200 nm pump pulse, is a rather slow, multi-picosecond process. More precisely, only one out of 18 initial conditions utilized in the EOM-CCSD simulations underwent internal conversion to S$_1$ in 500 fs, while $\sim 80\%$ of the population remains on S$_2$ after 1 ps in the TDDFT simulations. As a result, EOM-CCSD and TDDFT agree that the S$_2$ population decay time-constant is greater than 2 ps. While a more quantitative value would require longer dynamics simulations, these results nevertheless indicate that EOM-CCSD and TDDFT predict internal conversion out of S$_2$ to be a multi-picosecond process. The molecule relaxes to a minimum of Rydberg character on the S$_2$ potential energy surface following photoexcitation, and slow internal conversion can only happen through vibrations away from this minimum. The transient UED signal computed within the IAM (shown in Fig. \ref{fig:UED} from the AIMS simulations indeed shows both relaxation to the S$_2$ Rydberg geometry and subsequent $\sim$ 500 cm$^{-1}$ C-C ring stretching vibrations. We note that the differences in electron density between the ground state and the Rydberg excited state of this small molecule could lead to signal differences that would require more detailed UED simulations\citep{parrish2019aiscat} (not attempted here). Previous work has shown that beyond-IAM computations can be important in modeling diffraction.\citep{weber2020rydbergchd,champenois2023ammonia,yang2020simultaneous}

The small amount of internal conversion to S$_1$ which was observed in the TDDFT simulations arises from structural changes involving C-C and C-H bond elongations (as shown in Fig. \ref{fig:IC_analysis}). 
EOM-CCSD is expected to lead to even slower dynamics as the resulting S$_2\to$ S$_1$ CIs are generally more energetically inaccessible than TDDFT. Nonetheless, EOM-CCSD does appear to have a lower barrier for relaxation via C$_O-$C$_\alpha$ than TDDFT, suggesting that this pathway is more relevant than indicated by the TDDFT/AIMS internal conversion pathway analysis in Fig. \ref{fig:IC_analysis}.
We also note that the role of C-H elongations in internal conversion in particular could potentially be experimentally studied with time-resolved experiments on deuterated cyclobutanone.  
As discussed in the SI, we also find that intersystem crossing to the triplet manifold is unlikely on the few picosecond timescale. The spin-orbit couplings between the S$_2$ state at the Rydberg minimum and proximate triplets (T$_{\{2,3\}}$) is $< 1 $ cm$^{-1}$ (though larger values are potentially possible in the stretched bond regimes). 

Both TDDFT and EOM-CCSD perform linear-response on single-reference ground state solutions and therefore can have difficulty with molecular geometries involving dissociating bonds. Specifically, overestimation of ground state energies due to persistence of charge-transfer contributions in the reference within a spin-restricted formalism could lead to elevated excited state energies in the stretched bond regime, creating artificial barriers against dissociation\citep{hait2019beyond}. However, EOM-CCSD is expected to exhibit such behavior to a \textit{lesser} extent than TDDFT (as it explicitly correlates electrons in a pairwise manner), and the \textit{apparently slower} internal conversion dynamics in the former therefore appears to lend greater confidence to the few-picosecond internal conversion timescale predicted by TDDFT. This suggests that the largest disagreement between our computational results and the experiment would arise from the underlying electronic structure methods being challenged by bond cleavage, followed by the use of the IAM to compute UED signals.  In addition, we note that we sampled from the ground state harmonic vibrational wavefunction of the molecule, neglecting any anharmonic effects or finite-temperature contributions. Indeed, cyclobutanone has a low frequency ( $<100$ cm$^{-1}$) ring puckering mode with substantial anharmonic character\cite{tamagawa1983molecular}, and this degree of freedom may not be well approximated by our ground state harmonic wavefunction based approach to initial conditions for AIMS simulations.  

We have tried to stress the need for continual validation before, during, and after the dynamics simulations, through comparisons to more accurate electronic structure methods or available experiments. Indeed, the computation of dynamics with the EOM-CCSD method (using relatively few initial conditions due to the computational expense) may be viewed as an extreme example of such continual validation. In this case, the EOM-CCSD dynamics boosted our confidence in a multi-picosecond S$_2$ lifetime. However, we also encountered a warning in the last comparison (after the dynamics, comparing to a TRPES experiment). In particular, our simulations of the TRPES observable with the same dynamics we used for predicting the UED observable suggest that both EOM-CCSD and TDDFT may be over-predicting the lifetime of S$_2$. It will be exciting to see how well our predictions fare. As Yogi Berra and Niels Bohr famously said, "Prediction is hard, especially about the future."

\section*{Acknowledgment} 
This research was financially supported by the AMOS program within the U.S. Department of
Energy, Office of Science, Basic Energy Sciences, Chemical Sciences, Geosciences, and
Biosciences Division. D.H. was supported by a Stanford Science Fellowship. O.J.F. was supported by the U.S. Department of Energy, Office of Science, Office of Advanced Scientific Computing Research, Department of Energy Computational Science Graduate Fellowship under Award Number DE-SC0024386. P.A.U. acknowledges support by the National Science Foundation MPS- Ascend Postdoctoral Research Fellowship, under Grant No. 2213324. This work used computational resources of the National Energy Research Scientific Computing Center (NERSC), a U.S. Department of Energy Office of Science User Facility located at Lawrence Berkeley National Laboratory, operated under Contract No. DE-AC02-05CH11231 using NERSC award BES-ERCAP0024588. 

\section*{Author Contributions}
DH, DL, and OJF carried out EOM-CCSD and TDDFT dynamics. DH created the first draft and led manuscript preparation. ASPP carried out TRPES calculations. PAU, BR, and LL participated in analysis and calculations. YW supported the dynamics and electronic structure simulations. TJM procured funding, led the effort, participated in analysis and manuscript writing. All authors reviewed the manuscript.   

\section*{Data Availability}
The data that supports the findings of this study are available within the article and its supplementary material.

\section*{Supporting Information}
PDF: Details about sensitivity to basis sets and functionals. Further information about initial conditions, characterization of S$_2$ minimum, discussion of likelihood of intersystem crossing, and detailed UED spectra.

\section*{Conflicts of Interest}
T.J.M. is a co-founder of PetaChem, LLC.
\bibliography{references}
\end{document}


\maketitle
\tableofcontents

\section{Experimental Protocol}
The details of the experiment associated with the special issue are provided below, reproduced verbatim from the invitation to the Special Issue from the website of the Journal of Chemical Physics:

``The experiment will be performed at the SLAC Megaelectronvolt Ultrafast Electron Diffraction facility. A gas sample of about 1 mbar of cyclobutanone will be irradiated with 200 nm light ($\approx$80 fs cross-correlation) and electron diffraction images will be obtained with 150 fs time resolution (FWHM) and 0.6 Å spatial resolution (2$\pi$/Smax), with the scattering vector S ranging from 1-10 {\AA}. The experiment will be performed at a repetition rate of 360 Hz and the gas sample will be exchanged after each optical/electron pulse pair. Note that the excitation is believed to target a Rydberg (3s) excited state (i.e., n$\to$3s) and not the n$\to\pi^*$ state (280 nm) which is believed to be the lowest singlet excited state. This is because the oscillator strength of the  n$\to\pi^*$ state is too low and direct excitation to this state would likely lead to multiphoton transitions. The intensity of the 200 nm excitation light will be kept as low as experimentally feasible (5 $\mu$J) to excite \&symp;10\% of the molecules and avoid multiphoton excitation. The experiment will collect diffraction images for time delays from -1 ps to 10's of ps in variable step sizes. The immediate region around time zero (-200 fs to 200 fs) will be scanned with 30 fs stepsize. Longer positive delays will be scanned with step sizes up to several ps. "

\section{Comparison to CC3 at the Franck-Condon Geometry}

\begin{table}[htb!]
\begin{tabular}{llllll}
\hline
\multicolumn{2}{c}{LRC-$\omega$PBE} & \multicolumn{2}{c}{EOM-CCSD} & \multicolumn{2}{c}{EOM-CC3} \\ \hline
E            & f             & E            & f             & E           & f             \\ \hline
4.19         & 0.0000        & 4.31         & 0.0000        & 4.28        & 0.0000        \\
6.44         & 0.0374        & 6.40         & 0.0390        & 6.32        & 0.0391        \\
7.09         & 0.0007        & 7.07         & 0.0013        & 6.98        & 0.0007        \\
7.21         & 0.0005        & 7.16         & 0.0000        & 7.09        & 0.0000        \\
7.27         & 0.0002        & 7.23         & 0.0016        & 7.14        & 0.0021        \\ \hline
\end{tabular}
\caption{Comparison of excitation energies (E, in eV) and oscillator strengths (f) between  LRC-$\omega$PBE/TDA, EOM-CCSD and EOM-CC3. The LRC-$\omega$PBE calculation was done at the ground state LRC-$\omega$PBE optimized geometry, while both EOM-CCSD and EOM-CC3 utilized the CCSD optimized ground state geometry. All calculations utilized the aug-cc-pVDZ basis. Only the results for the first five singlet excited states (n$\to\pi^*$, n$\to$3s and $n\to$3p by ascending order of energy) is shown}
\label{tab:cc3}
\end{table}
We assessed the performance of TDDFT and EOM-CCSD towards the prediction of the absorption spectrum through comparison with EOM-CC3\citep{koch1997cc3}, which is a higher level of coupled cluster theory that includes an approximate iterative description of triple excitations. Indeed, EOM-CC3 is found to significantly exceed the accuracy of EOM-CCSD in predicting vertical excitation energies of small molecules\citep{loos2018mountaineering}.  Comparing the excitation energies from the electronic structure methods (as shown in Table \ref{tab:cc3}), we find that EOM-CCSD and TDDFT slightly overestimate the excitation energy of the S$_2$ state at the Franck-Condon geometry. The agreement between EOM-CCSD and EOM-CC3 is however perceptibly greater, consistent with our expectation of EOM-CCSD being more accurate than TDDFT.

\section{Basis Set Dependence of LRC-$\omega$PBE Excitation Energies at Franck-Condon Geometry}
\begin{table}[htb!]
\begin{tabular}{llllllllllll}
\hline
\multicolumn{2}{c}{aug-cc-pVDZ} & \multicolumn{2}{c}{daug-cc-pVDZ} & \multicolumn{2}{c}{daug-cc-pVTZ} & \multicolumn{2}{c}{6-31+G**} & \multicolumn{2}{c}{aug-pc-1} & \multicolumn{2}{c}{def2-SVPD} \\ \hline
E             & f               & E              & f               & E              & f               & E            & f             & E            & f             & E            & f              \\
4.19          & 0.0000          & 4.19           & 0.0000          & 4.21           & 0.0000          & 4.25         & 0.0000        & 4.23         & 0.0000        & 4.21         & 0.0000         \\
6.44          & 0.0374          & 6.38           & 0.0339          & 6.43           & 0.0335          & 6.85         & 0.0451        & 6.46         & 0.0348        & 7.05         & 0.0399         \\
7.09          & 0.0007          & 6.98           & 0.0004          & 7.02           & 0.0004          & 7.54         & 0.0185        & 7.09         & 0.0006        &              &                \\
7.21          & 0.0005          & 7.06           & 0.0009          & 7.11           & 0.0009          & 7.69         & 0.0000        & 7.20         & 0.0012        &              &                \\
7.27          & 0.0002          & 7.14           & 0.0007          & 7.18           & 0.0008          & 7.71         & 0.0030        & 7.27         & 0.0002        &              &                \\ \hline
\end{tabular}
\caption{Comparison of excitation energies (E, in eV) and oscillator strengths (f) with LRC-$\omega$PBE/TDA, utilizing a number of standard basis sets at the Franck-Condon geometry optimized with aug-cc-pVDZ. Only the results for the first five singlet excited states (n$\to\pi^*$, n$\to$3s and $n\to$3p by ascending order of energy) is shown, except for def2-SVPD for which the 3p Rydberg states are greatly elevated in energy and could not be completely disambiguated from valence excitations. Terachem presently lacks f-orbital support and so the daug-cc-pVTZ calculation was performed with Q-Chem (with the two packages agreeing with each other for excitation energy predictions to 0.01 eV and oscillator strengths to 0.0001, with the aug-cc-pVDZ basis). }
\label{tab:daug}
\end{table}
The excitation energies of Rydberg states tend to be basis set sensitive and often require a large number of diffuse functions to converge\citep{liang2022revisiting}. We employed the aug-cc-pVDZ basis for our calculations, as it appeared to represent a reasonable balance between low basis set incompleteness error and computational ease (highly diffuse basis sets being computationally challenging not only on account of increased number of basis functions but also linear dependencies between them). Comparison to the doubly augmented daug-cc-pVDZ basis\cite{woon1994gaussian} indicates that the S$_2$ excitation energy can be lowered by 0.06 eV in the presence of additional diffuse functions (see Table \ref{tab:daug}). However, the larger daug-cc-pVTZ basis yields slightly higher excitation energies than duag-cc-pVDZ (6.43 eV vs 6.38 eV), which is in better agreement with the aug-cc-pVDZ excitation energy (6.44 eV). For practical purposes therefore, aug-cc-pVDZ appears to be adequate for representing the S$_2$ character Rydberg excitation. 

We also compare aug-cc-pVDZ results to other widely used basis set families of comparable (augmented double zeta) size in Table \ref{tab:daug}. The Jensen polarization consistent aug-pc-1 basis\cite{jensen2001polarization,jensen2002polarization,jensen2002polarizationb} yields Rydberg excitation energies that are fairly close to aug-cc-pVDZ.  However the popular def2-SVPD basis\cite{weigend2005balanced,rappoport2010property} elevates the energy of the S$_2$ state by 0.6 eV, and the 3p Rydberg states cannot even be readily identified as they are pushed inside the high energy valence manifold. This result is consistent with earlier reports of rather poor performance of the def2 basis sets for modeling Rydberg states of small molecules\cite{liang2022revisiting}. The Pople 6-31+G* represents a more intermediate case in which the Rydberg excitations energies is considerably elevated (by 0.4-0.5 eV) relative to aug-cc-pVDZ but to a lesser extent than def2-SVPD. We therefore recommend that careful basis set benchmarking be carried out before utilization of Pople or def2 basis sets for modeling of Rydberg states of small molecules. We do also note that all basis sets tested in Table \ref{tab:daug} yield quite similar S$_1$ excitation energies (4.19-4.25 eV), indicating the lack of strong basis set effects for modeling valence excitations with LRC-$\omega$PBE.

\section{Functional Dependence of aug-cc-pVDZ Excitation Energies at Franck-Condon Geometry}
\begin{table}[htb!]
\begin{tabular}{llllllllll}
\hline
\multicolumn{2}{c}{LRC-$\omega$PBE} & \multicolumn{2}{c}{B3LYP} & \multicolumn{2}{c}{PBE} & \multicolumn{2}{c}{PBE0} & \multicolumn{2}{c}{LRC-$\omega$PBEh} \\ \hline
E            & f             & E          & f            & E         & f           & E          & f           & E            & f              \\
4.19         & 0.0000             & 4.20       & 0.0000            & 4.02      & 0.0000           & 4.24       & 0.0000           & 4.26         & 0.0000              \\
6.44         & 0.0374        & 5.81       & 0.0411       & 5.07      & 0.0363      & 6.08       & 0.0417      & 6.35         & 0.0388         \\
7.09         & 0.0007        & 6.42       & 0.001        & 5.72      & 0.0038      & 6.68       & 0.0004      & 6.96         & 0.0005         \\
7.21         & 0.0005        & 6.52       & 0.0045       & 5.80      & 0.0074      & 6.79       & 0.0033      & 7.09         & 0.0001         \\
7.27         & 0.0002        & 6.54       & 0.001        & 5.84      & 0.0016      & 6.79       & 0.0007      & 7.12         & 0.0001         \\ \hline
\end{tabular}
\caption{Comparison of excitation energies (E, in eV) and oscillator strengths (f) with a number of different functionals and the aug-cc-pVDZ basis at the Franck-Condon geometry optimized with LRC-$\omega$PBE/aug-cc-pVDZ. Only the results for the first five singlet excited states (n$\to\pi^*$, n$\to$3s and $n\to$3p by ascending order of energy) is shown. The experimental band maximum for the S$_2$ state is at $\sim$6.40 eV.}
\label{tab:functional}
\end{table}

The excitation energies of Rydberg states in TDDFT calculations are also typically quite sensitive to the choice of density functional\cite{tozer1998improving}, roughly increasing monotonically with the proportion of Hartree-Fock (HF) exchange present. This can be understood by viewing a Rydberg excitation as a long-range charge-transfer like process where the electron is excited to a diffuse level with very low spatial overlap with the hole formed in the valence region, which semilocal density functionals cannot qualitatively describe within a linear-response formalism\cite{dreuw2005single}. HF in principle can describe the particle-hole interaction correctly in the asymptotic limit of large separation between them, but the lack of correlation leads to a significant overestimation of Rydberg excitation energies with pure TDHF. Long range corrected hybrid functionals (where the HF exchange contribution rises to $100\%$ at infinite separation from a lower value at shorter separation) therefore represent the optimal route for modeling Rydberg excitations within TDDFT\cite{liang2022revisiting}.  

We compared the performance of a number of popular density functional approximations towards prediction of the S$_2$ state excitation energy from TDDFT, with the results being shown in Table \ref{tab:functional}. We find that LRC-$\omega$PBE is quite effective at reproducing the experimental band maximum ($\sim$ 6.4 eV), but the other tested functionals do not fare as well. The local PBE\cite{PBE} functional underestimates the Rydberg state energy by about $\sim 1.4$ eV, while the popular global hybrid functionals B3LYP\cite{b3lyp} and PBE0\cite{pbe0} underestimate by $0.6$ and $0.4$ eV respectively. Interestingly, the related LRC-$\omega$PBEh\cite{lrcwpbeh} functional slightly underestimates the S$_2$ excitation energy as well, likely due to the smaller range separation parameter $\omega$ (0.2 a.u. vs 0.3 a.u. for LRC-$\omega$PBE). 

We note that it is possible to use functional/basis set combinations that reasonably reproduce the experimental S$_2$ excitation energy via cancellation of errors (i.e. overestimation arising from basis set incompleteness compensated by underestimation arising from semilocal exchange-correlation approximation), but it is harder to predict if such compensation would hold at other geometries which may be accessed through the course of dynamics simulations. We have therefore attempted to minimize basis set incompleteness error through the use of aug-cc-pVDZ, and utilized a functional (LRC-$\omega$PBE) that predicts the experimental absorption spectrum well with this basis.

\section{Analysis of TDDFT/AIMS Initial Conditions}
\begin{figure}[!hb]
    \includegraphics[width=0.7\linewidth]{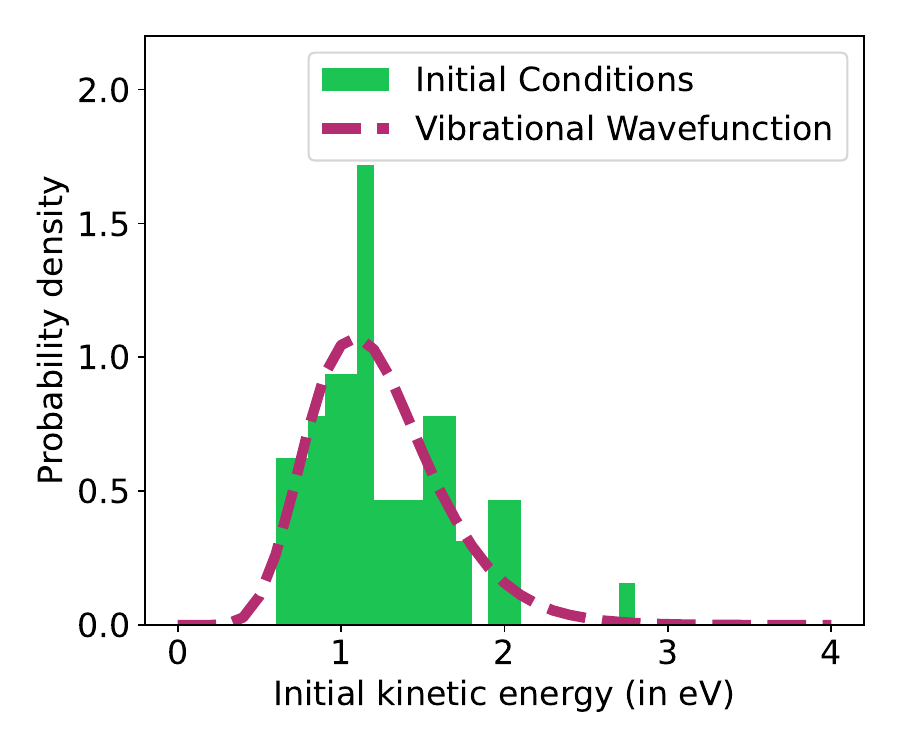}
\caption{Classical nuclear kinetic energy distribution of the 64 TDDFT initial conditions (histogram, with 0.1 eV bin width), compared to the distribution arising from the ground state vibrational wavefunction within the harmonic approximation. The initial condition average is 1.25 eV, with a standard deviation of 0.43 eV. This is quite close to the average of 1.23 eV and a standard deviation of 0.40 eV from the classical nuclear kinetic energy distribution arising from the ground state vibrational wavefunction. The latter distribution is constructed from a Gaussian kernel density estimation on a million samples drawn from the ground state vibrational wavefunction for the construction of this distribution alone (and not utilized elsewhere). }
\label{fig:KEcomp}
\end{figure}
The Wigner phase-space quasiprobability distribution\cite{wigner1932quantum} corresponding to the ground state harmonic oscillator wavefunction is completely separable into independent distributions for nuclear positions and momenta. The selection of nuclear positions for the AIMS simulations is described in detail in the main text, but it is worth noting that the sampling protocol involved additional weighting for S$_2$ excitation energies and oscillator strengths (on top of sampling from the ground state harmonic position distribution) to better account for the 200 nm pump pulse. Indeed, the selected initial positions have S$_2$ excitation energies centered about 200 nm, as shown in Fig 2. of the main text. 

No such weighting for the pump is required for the initial nuclear momentum sampling, and the momenta were thus directly drawn from the distribution arising from the corresponding ground state harmonic wavefunction. The finite number of initial momenta sampled for dynamics (64 for TDDFT and 18 for EOM-CCSD) could potentially lead to a scenario in which relatively low probability outliers are over-represented in the initial conditions, affecting the statistical averages over the AIMS simulations. As a very simple test for this, we computed the classical nuclear kinetic energy distribution arising from both the 64 TDDFT initial conditions, and the true distribution expected from the ground state harmonic wavefunction (here approximated computationally via Gaussian kernel density estimation on a million samples drawn from the ground state vibrational wavefunction solely for the purpose of constructing this distribution and not for any other application). The comparison (shown in Fig. \ref{fig:KEcomp}) between the histogram arising from the 64 initial nuclear momenta used for the TDDFT/AIMS dynamics with the continuous probability density distribution expected over the ground state harmonic wavefunction show a reasonable level of agreement, though it can be potentially argued that somewhat higher nuclear kinetic energies were slightly over-represented in the initial conditions for our TDDFT/AIMS simulations.   

\section{Comment on TDDFT/AIMS Calculations}
The computational resources required by an AIMS simulation is inversely related to the rate of internal conversion, as the latter entails spawning events that lead to formation of new trajectories whose energies and forces also need to be computed. In general, the per-timestep cost of an AIMS simulation involving $N$ trajectories costs somewhere between $N$ and $\dfrac{N(N+1)}{2}$ times the cost of an individual trajectory (as centroids between trajectories also need to be simulated in the regime where trajectory basis functions show substantial overlap\citep{curchod2018ab}). One of the TDDFT/AIMS initial conditions led to substantial spawning rather early on in time, leading to a large number of trajectories that prevented us from being able to run this initial condition to 1 ps. Consequently, the results presented in the main text only utilize data from the simulations starting from the other 63 TDDFT/AIMS initial conditions, unless specified otherwise. We do note that 88\% of the S$_2$ population corresponding to this initial condition was transferred to S$_1$ within 379 fs (the last timestep in the AIMS simulation that this initial condition was run to), indicating that complete internal conversion to S$_1$ within 1 ps is a not unreasonable assumption for this trajectory.

\newpage 
\section{Normal modes of the Rydberg S$_2$ minimum}
\begin{table}[htb!]
\begin{tabular}{lll}
\hline
\multicolumn{1}{c}{Mode number} & \multicolumn{1}{c}{Frequency (cm$^{-1}$)} & \multicolumn{1}{c}{Description} \\ \hline
1 & 146.827 & Pucker \\ 
2 & 324.482 & Carbonyl in-plane bend \\ 
3 & 478.687 & C-C, C-H stretch\\ 
4 & 510.499 & C-C stretch \\ 
5 & 514.722 & C-C stretch \\ 
6 & 536.341 & C-C stretch \\ 
7 & 704.473 & C-C, C-H stretch \\ 
8 & 786.285 & C-C stretch \\ 
9 & 964.581 & Hydrogen twist \\ 
10 & 999.596 & C-C stretch \\ 
11 & 1009.150 & Hydrogen twist \\ 
12 & 1018.546 & C-C stretch \\ 
13 & 1139.139 & Hydrogen wag \\ 
14 & 1192.683 & Hydrogen twist \\ 
15 & 1214.451 & Hydrogen wag \\ 
16 & 1220.565 & Hydrogen twist \\ 
17 & 1251.600 & Hydrogen wag \\ 
18 & 1279.785 & Hydrogen scissor \\ 
19 & 1288.284 & Hydrogen scissor \\ 
20 & 1421.184 & Hydrogen scissor \\ 
21 & 1954.035 & Carbonyl stretch \\ 
22 & 2905.244 & Symmetric hydrogen stretch \\ 
23 & 3045.425 & Asymmetric hydrogen stretch \\ 
24 & 3106.240 & Symmetric hydrogen stretch \\ 
25 & 3179.695 & Asymmetric hydrogen stretch \\ 
26 & 3240.777 & Asymmetric hydrogen stretch \\ 
27 & 3552.799 & Symmetric hydrogen stretch \\ 

\end{tabular}
\caption{Normal mode indices, frequencies, and descriptions for the Rydberg S$_2$ minimum, as obtained from LRC-$\omega$PBE/aug-cc-pVDZ with TDDFT. No imaginary frequencies were found, confirming that the critical point is indeed a minimum.}
\label{tab:s2_modes}
\end{table}

\section{Discussion on Potential Intersystem Crossing}
\begin{table}[htb!]
\begin{tabular}{lllll}
\hline
   & \multicolumn{2}{c}{LRC-$\omega$PBE}                   & \multicolumn{2}{c}{EOM-CCSD}                \\ \hline 
   & E (in eV) & SOC (in cm$^{-1}$) & E (in eV) & SOC (in  cm$^{-1}$) \\ \hline 
T$_1$ & -2.19   & 1.095647                        & -1.65   & 0.894605                        \\
T$_2$ & -0.14   & 0.000421                        & -0.10   & 0.000113                        \\
T$_3$ & 0.61    & 0.158047                        & 0.69    & 0.095279                        \\
T$_4$ & 0.75    & 0.771982                        & 0.70    & 0.001321                        \\
T$_5$ & 0.77    & 0.019405                        & 0.77    & 0.674356   \\     \hline              
\end{tabular}
\caption{Energies of triplet states (relative to S$_2$) and spin-orbit coupling (SOC) between these states and S$_2$ (evaluated using a mean-field approach for the two electron contributions\citep{kotaru2022spin,pokhilko2019general}) at the Rydberg type minimum optimized with the respective electronic structure method. These calculations were done with the Q-Chem software package\citep{epifanovsky2021software}.}
\label{tab:soc}
\end{table}
Our computational prediction of a long lived S$_2$ state from TDDFT/AIMS and EOM-CCSD/AIMS simulations only considered internal conversion to other singlet states. In principle, intersystem crossing to the triplet manifold is also a possible relaxation mechanism, although it is expected to be slow for a system like cyclobutanone where no heavy atoms are present. 

We attempt to approximate the rate of intersystem crossing out of S$_2$ by considering the spin-orbit coupling (SOC) matrix elements between the S$_2$ state at the Rydberg minimum geometry and triplet states at the TDDFT and EOM-CCSD level, which are shown in Table \ref{tab:soc}.  These SOC elements are rather small relative to the energy gaps between S$_2$ and the triplet states at this geometry. Consequently, intersystem crossing does not appear to be a significant factor in the few-picosecond photorelaxation process out of the S$_2$ Rydberg minimum. Triplet formation may however be more significant in the stretched C-C bond regimes associated with internal conversion (due to the presence of multiple low energy valence electronic states at such geometries) and thereby affect structural dynamics. 

\section{Momentum space diffraction patterns and further analysis of UED}

\begin{figure}[htb!]
\begin{minipage}{0.60\textwidth}
    \centering
    \includegraphics[width=\linewidth]{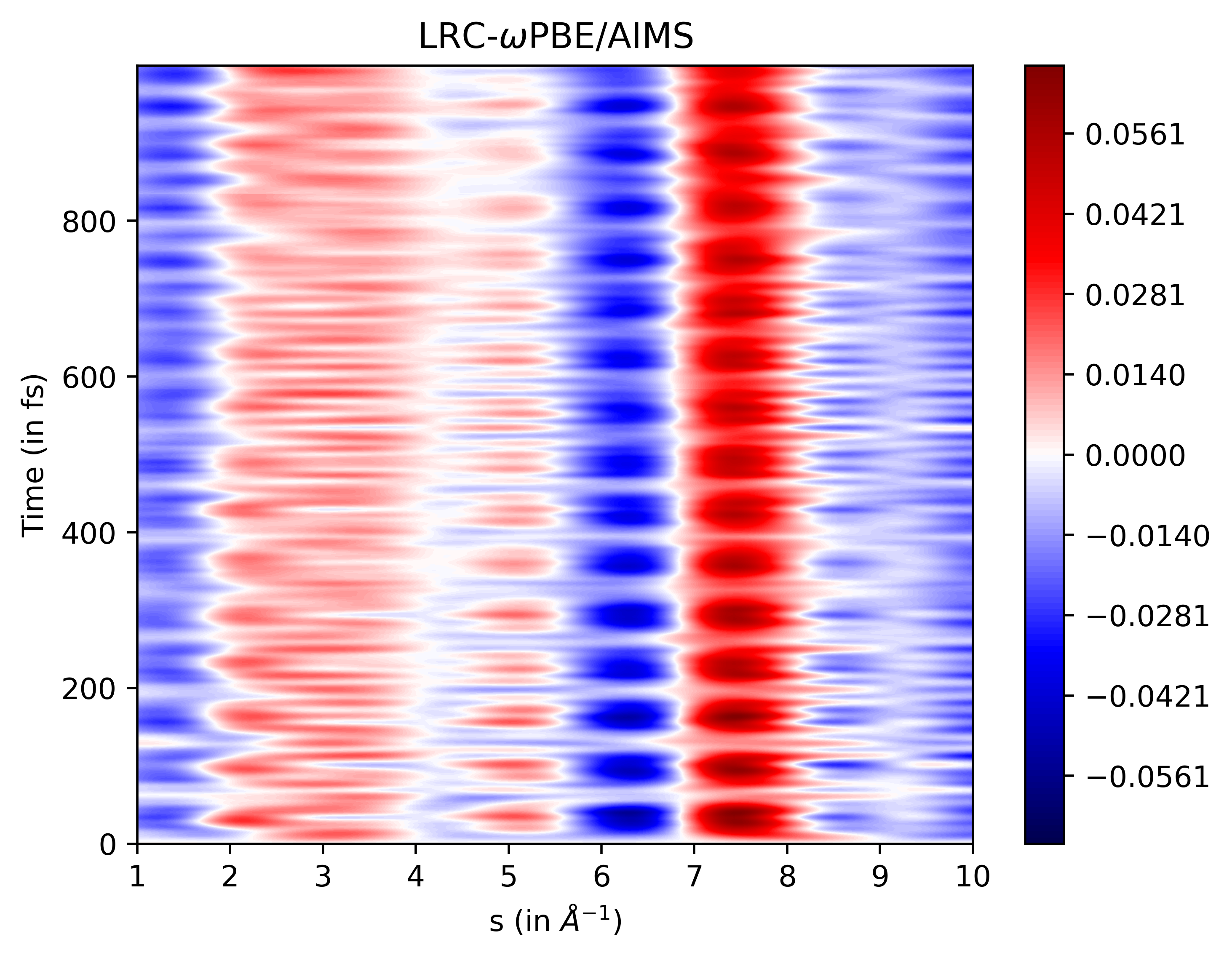}
\end{minipage}
\caption{Momentum space $\Delta I/I_0$ UED spectrum for LRC-$\omega$PBE/AIMS simulations. No smoothing applied in the time domain.}
\label{fig:TDDFT_kspace}
\end{figure}

\begin{figure}[htb!]
\begin{minipage}{0.60\textwidth}
    \centering
    \includegraphics[width=\linewidth]{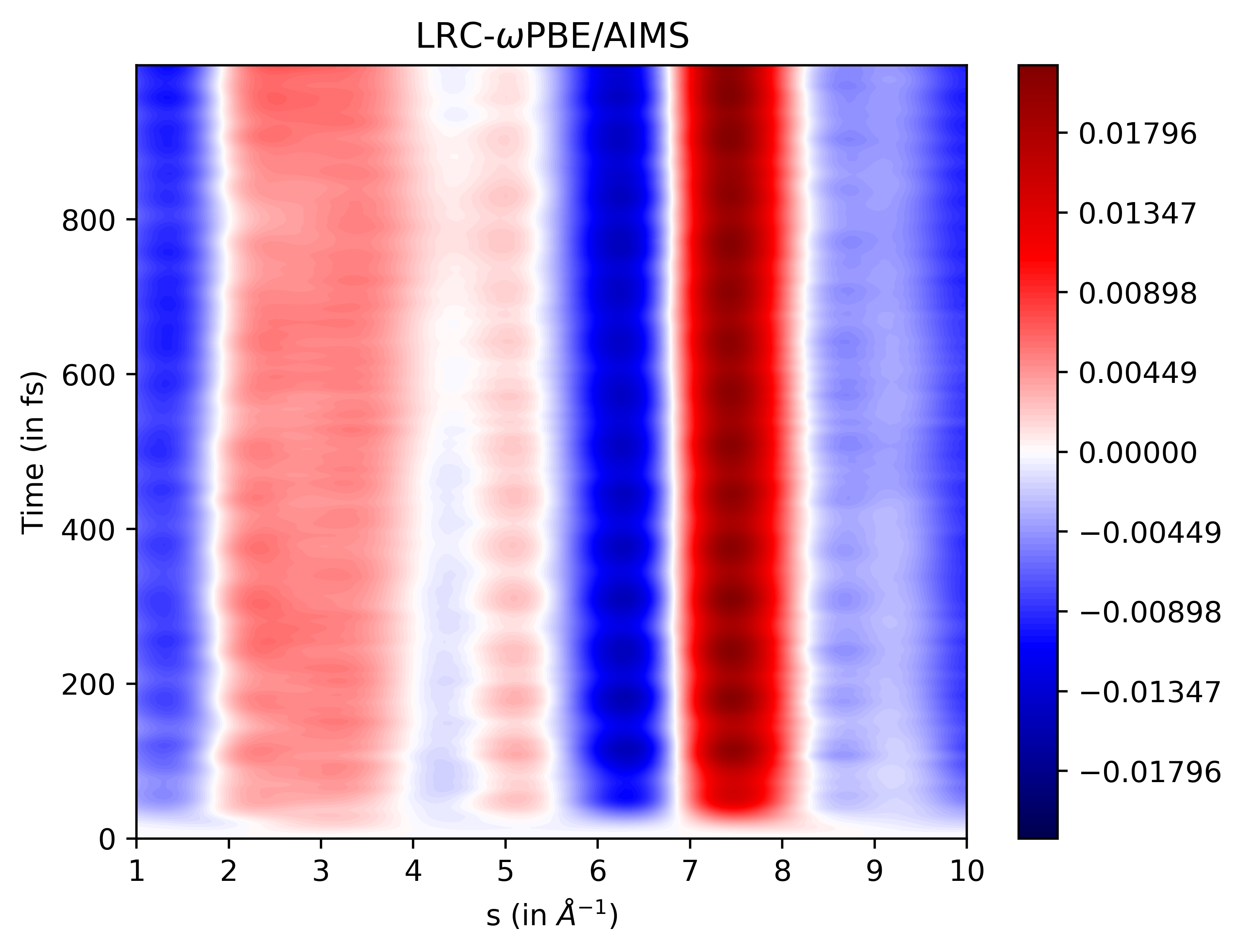}
\end{minipage}
\caption{Momentum space $\Delta I/I_0$ UED spectrum for LRC-$\omega$PBE/AIMS simulations. 150 fs FWHM smoothing applied in the time domain.}
\label{fig:TDDFT_kspace_convolved}
\end{figure}

\begin{figure}[htb!]
\begin{minipage}{0.60\textwidth}
    \centering
    \includegraphics[width=\linewidth]{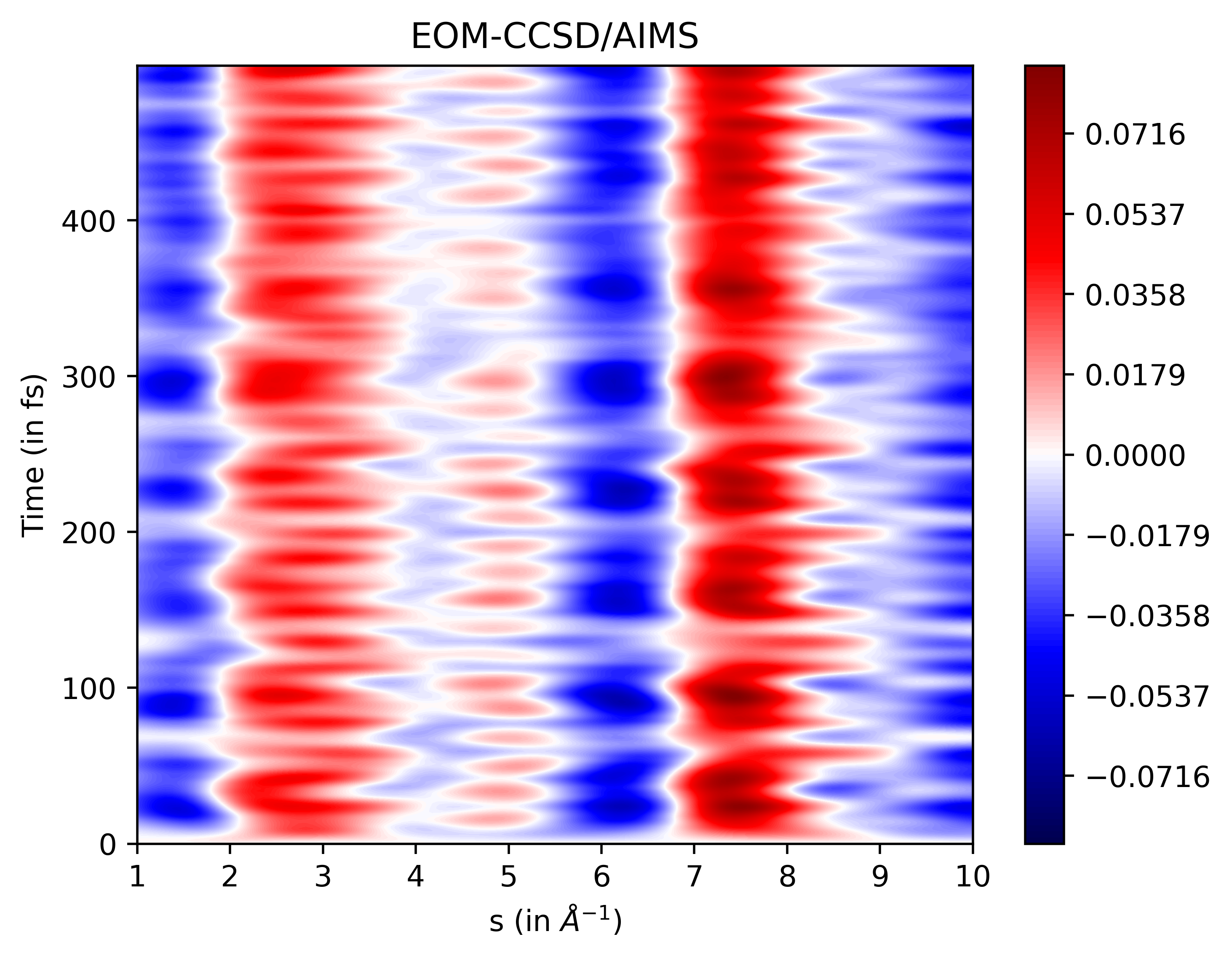}
\end{minipage}
\caption{Momentum space $\Delta I/I_0$ UED spectrum for EOM-CCSD/AIMS simulations. No smoothing applied in the time domain.}
\label{fig:EOMCCSD_kspace}
\end{figure}

\begin{figure}[htb!]
\begin{minipage}{0.60\textwidth}
    \centering
    \includegraphics[width=\linewidth]{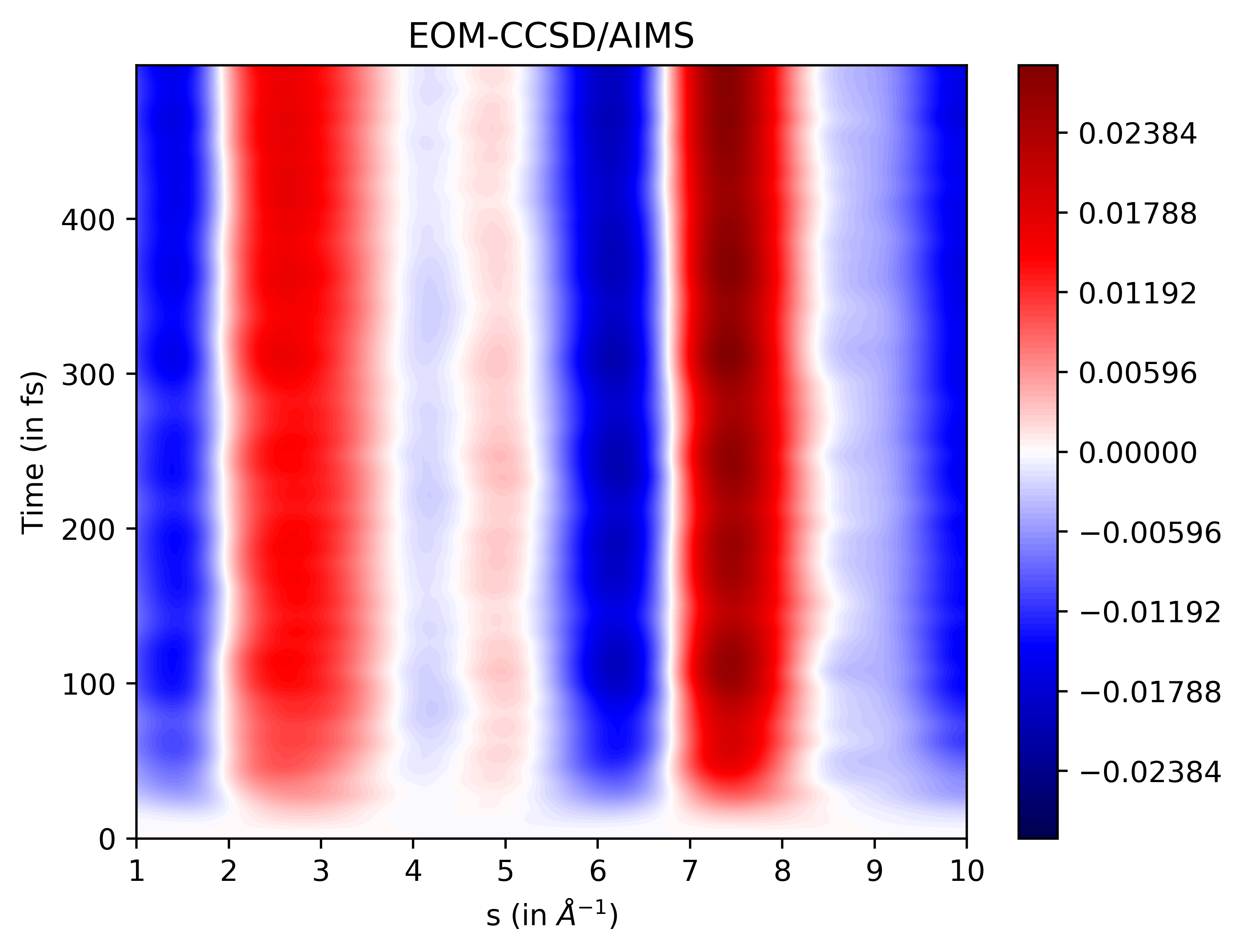}
\end{minipage}
\caption{Momentum space $\Delta I/I_0$ UED spectrum for EOM-CCSD/AIMS simulations. 150 fs FWHM smoothing applied in the time domain.}
\label{fig:EOMCCSD_kspace_convolved}
\end{figure}

\begin{figure}[htb!]
\begin{minipage}{0.60\textwidth}
    \centering
    \includegraphics[width=\linewidth]{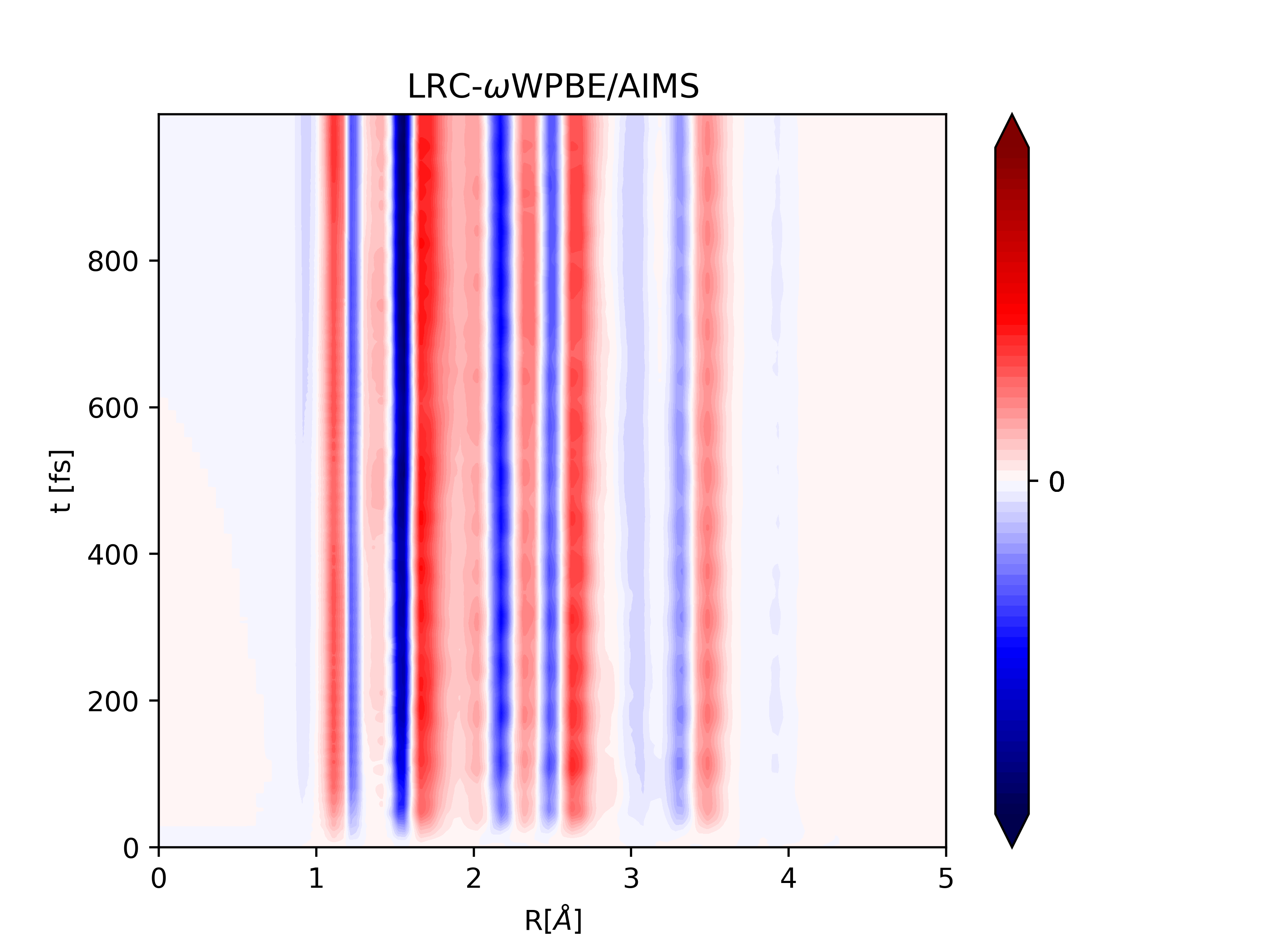}
\end{minipage}
\caption{Real space $\Delta I/I_0$ UED spectrum for LRC-$\omega$PBE/AIMS simulations. 150 fs FWHM smoothing applied in the time domain, and 0.1 {\AA} FWHM smoothing applied in the distance domain.}
\label{fig:TDDFT_01eV_FWHM}
\end{figure}

\begin{figure}[htb!]
\begin{minipage}{0.60\textwidth}
    \centering
    \includegraphics[width=\linewidth]{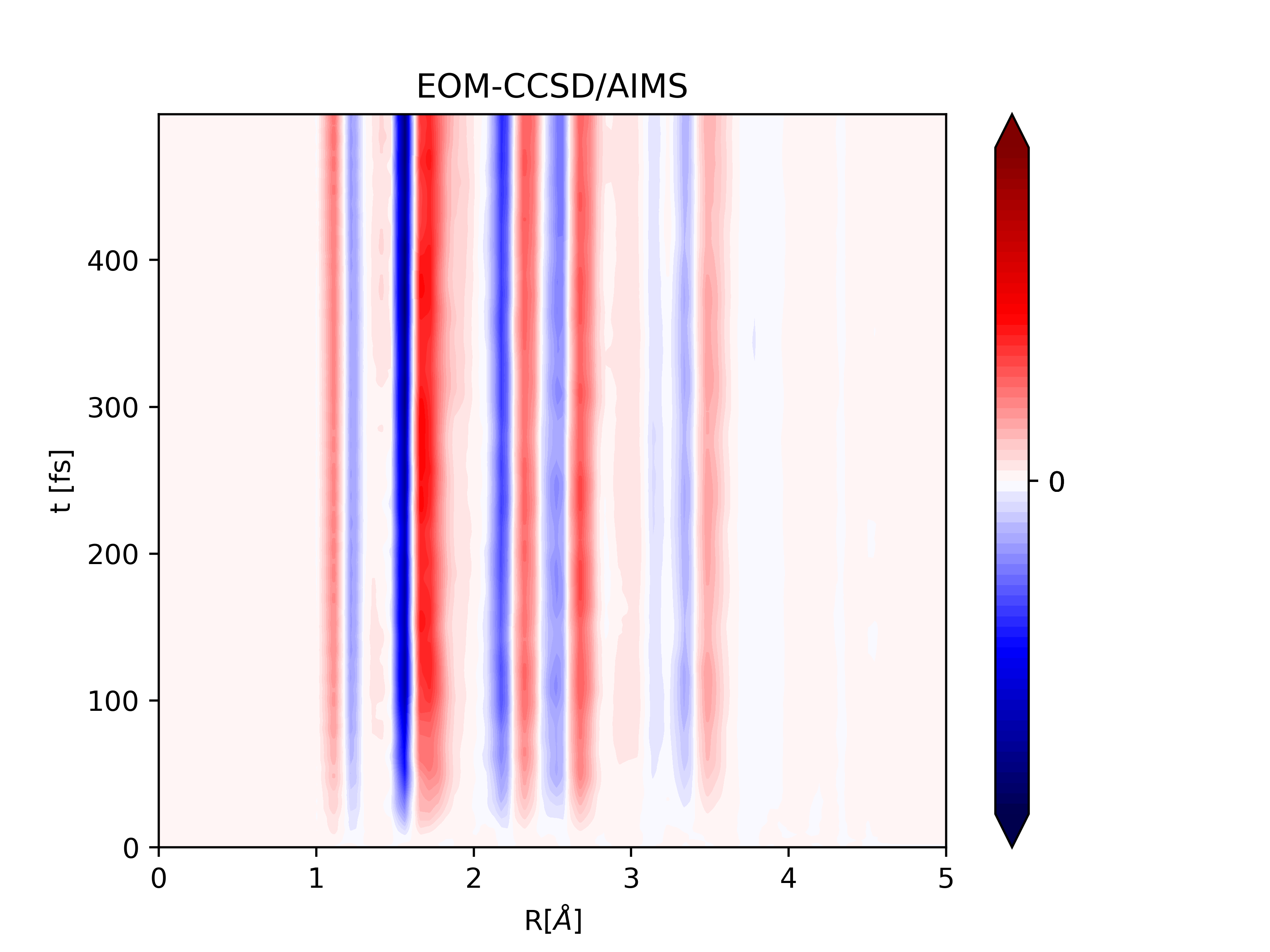}
\end{minipage}
\caption{Real space $\Delta I/I_0$ UED spectrum for EOM-CCSD/AIMS simulations. 150 fs FWHM smoothing applied in the time domain, and 0.1 {\AA} FWHM smoothing applied in the distance domain.}
\label{fig:TDDFT_01eV_FWHM}
\end{figure}

\begin{figure}[htb!]
\begin{minipage}{0.90\textwidth}
    \centering
    \includegraphics[width=\linewidth]{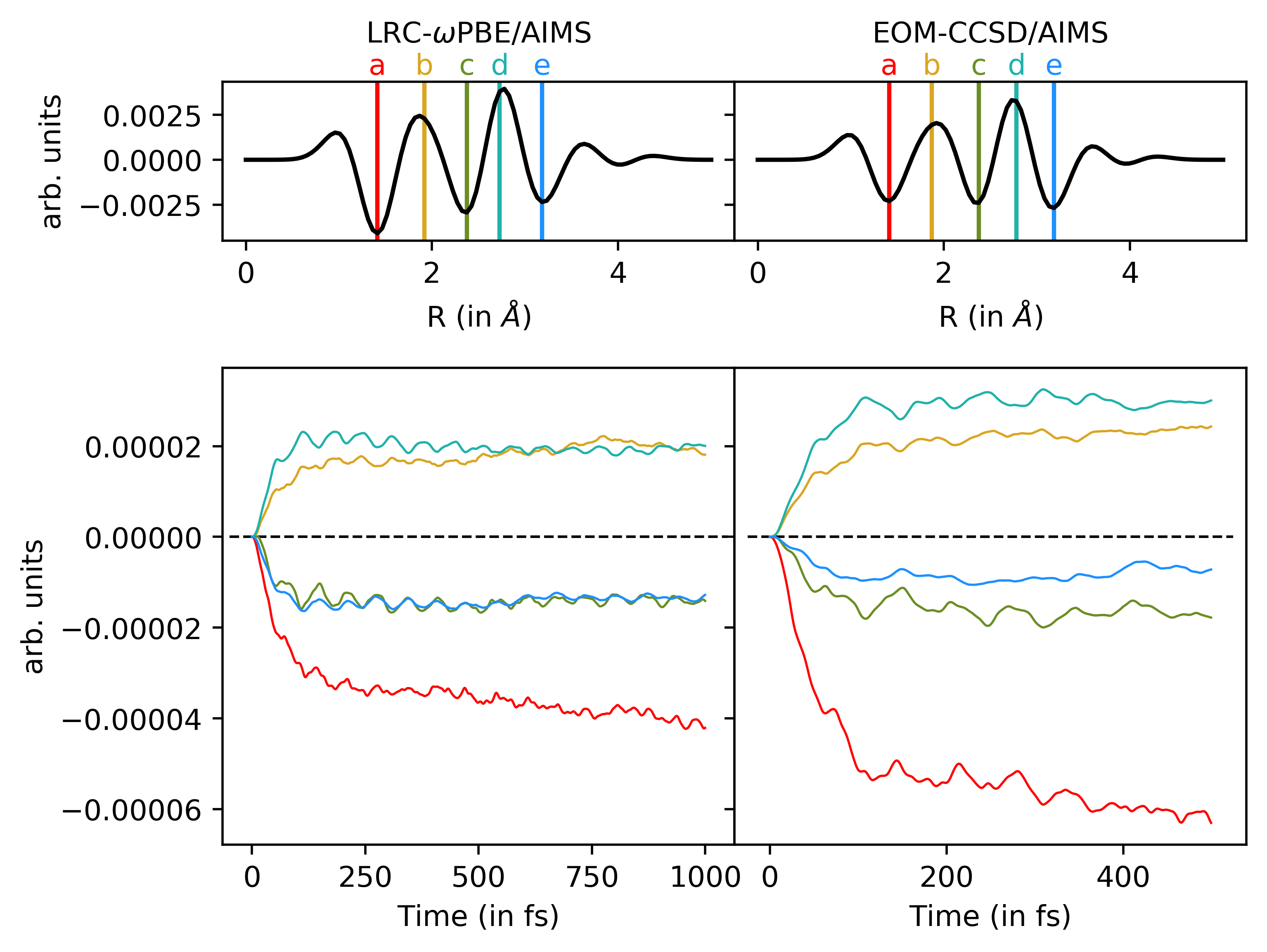}
\end{minipage}
\caption{Top panel: Real space $\Delta I/I_0$ signals generated from the $S_0$ min and $S_2$ Rydberg minimum optimized at the LRC-$\omega$PBE and EOM-CCSD levels of theory, respectively. a, b, c, d, and e label significant accumulations and depletions that persist in the time-resolved real space UED spectra. Bottom panel: Line-outs of the time-convolved real space UED for the $R$ values corresponding to a, b, c, d, and e from the top panel. All line-outs show a rapid growth/decay ($\sim$100 fs), followed by oscillatory behavior.}
\label{fig:UED_lineouts}
\end{figure}

\begin{figure}[htb!]
\begin{minipage}{0.90\textwidth}
    \centering
    \includegraphics[width=\linewidth]{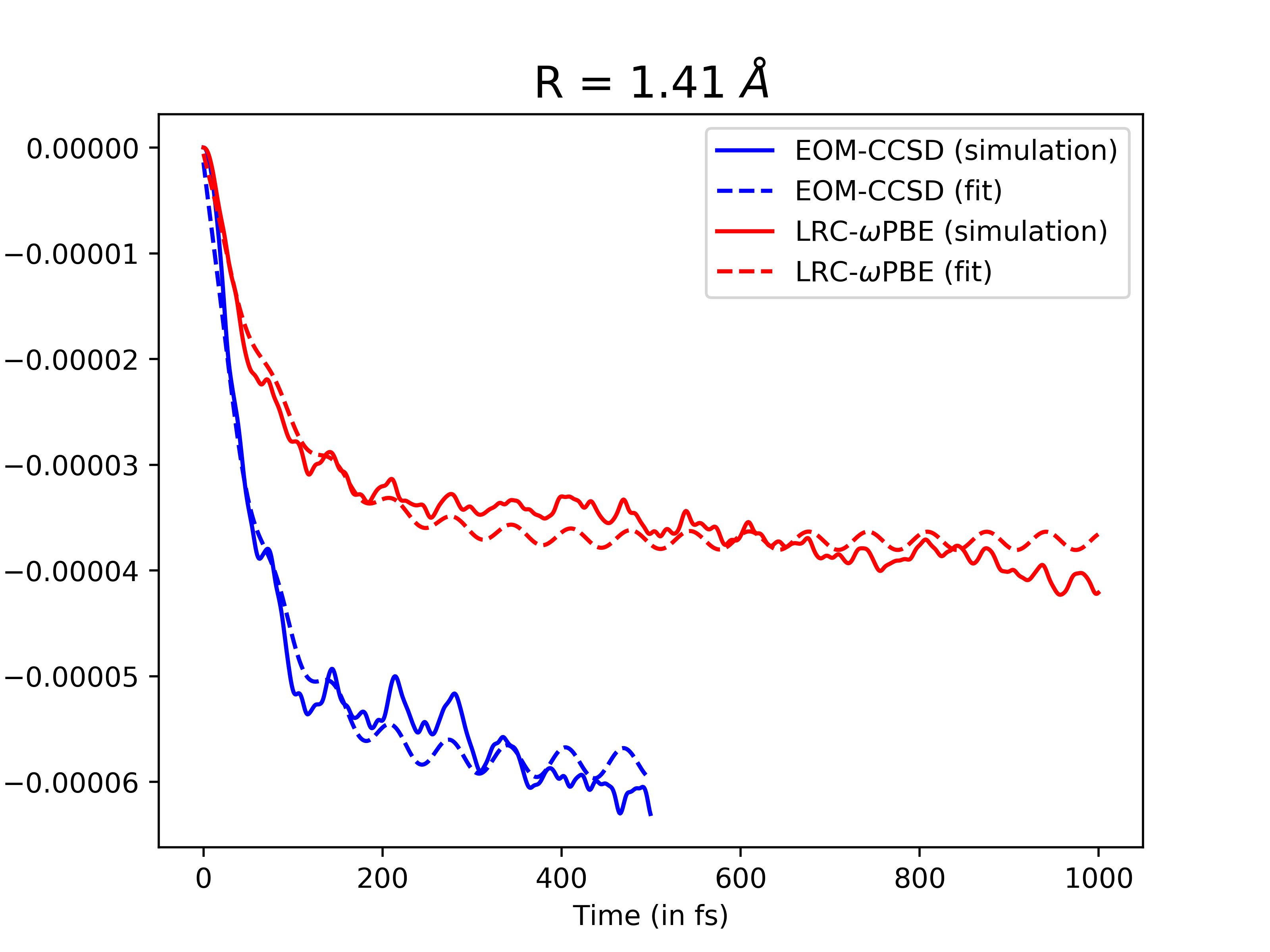}
\end{minipage}
\caption{Lineouts of a depletion feature in the time convolved UED at 1.41 {\AA} for TDDFT (red) and EOM-CCSD (blue).  The lineout is fit to $A (1 - c_1 e^{-t/\tau}- (1-c_1) \textrm{cos}(\omega t + \phi))$, where $\tau$ is the decay constant in fs, $\omega$ is the angular frequency of oscillation in fs$^{-1}$, $\phi$ is a phase shift, $c_1$ controls the relative influence of the two terms, and $A$ scales the amplitude. The fit produced $A$ = -5.827e-5, $c_1$ = 0.9753, $\tau$ = 1/65.14 fs, $\omega$ = 64.75 fs (or 515.2 cm$^{-1}$), and $\phi$ = -1.544 rad for EOM-CCSD and $A$ = -3.719e-5, $c_1$ = 0.9769, $\tau$ = 1/86.23 fs, $\omega$ = 66.29 fs (or 503.2 cm$^{-1}$), and $\phi$ = -1.243 rad for TD-DFT.}
\label{fig:bleach_fit}
\end{figure}

\begin{figure}[htb!]
\begin{minipage}{0.90\textwidth}
    \centering
    \includegraphics[width=\linewidth]{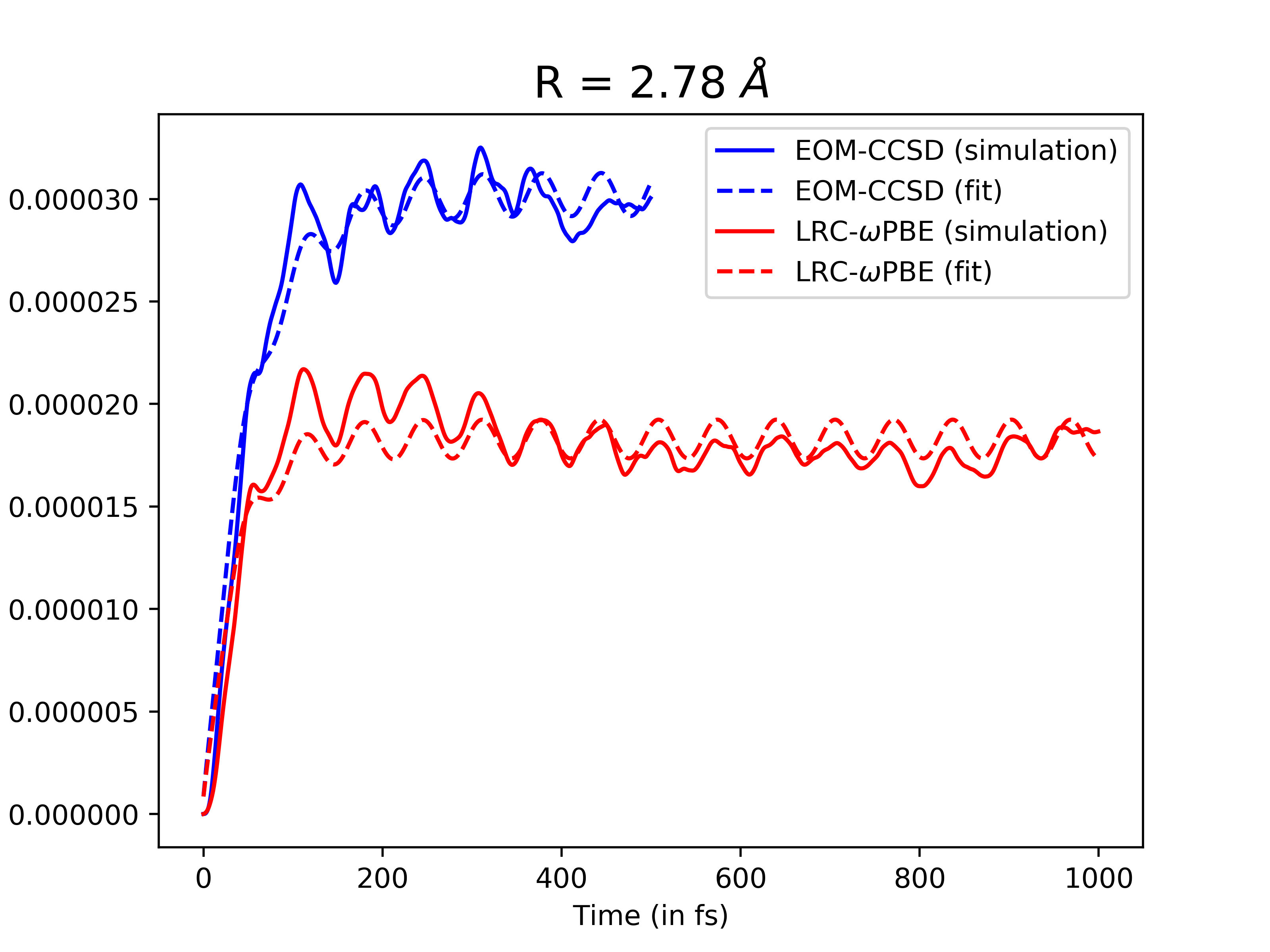}
\end{minipage}
\caption{Lineouts of a growth feature in the time convolved UED at 2.78 {\AA} for TDDFT (red) and EOM-CCSD (blue).  The lineout is fit to $A (1 - c_1 e^{-t/\tau}- (1-c_1) \textrm{cos}(\omega t + \phi))$, where $\tau$ is the decay constant in fs, $\omega$ is the angular frequency of oscillation in fs$^{-1}$, $\phi$ is a phase shift, $c_1$ controls the relative influence of the two terms, and $A$ scales the amplitude. The fit produced $A$ = 3.022e-5, $c_1$ = 0.9653, $\tau$ = 1/51.24 fs, $\omega$ = 66.08 fs (or 504.8 cm$^{-1}$), and $\phi$ = -1.392 rad for EOM-CCSD and $A$ = 1.829e-5, $c_1$ = 0.9486, $\tau$ = 1/36.00 fs, $\omega$ = 65.72 fs (or 507.6 cm$^{-1}$), and $\phi$ = -1.470 rad for TD-DFT.}
\label{fig:growth_fit}
\end{figure}

Momentum space $\Delta I/I_0$ UED spectra of the LRC-$\omega$PBE/AIMS dynamics are shown with and without 150 fs FWHM gaussian time convolution in Figures \ref{fig:TDDFT_kspace_convolved} and \ref{fig:TDDFT_kspace}, respectively. Likewise,  spectra of the EOM-CCSD/AIMS dynamics are shown with and without 150 fs FWHM gaussian time convolution in Figures \ref{fig:EOMCCSD_kspace_convolved} and \ref{fig:EOMCCSD_kspace}, respectively. These are provided for purposes of comparison to the experiment. 

We also provide real space spectra produced with a 0.1 {\AA} spatial convolution. The narrower spatial resolution enables the identification of more features, but these features will ultimately wash out in the real experimental signal due to the spatial resolution of the experiment.

\clearpage

\section{Gaussian Widths of Trajectory Basis Functions}
\begin{table}[htb!]
    \centering
    
    \begin{tabular}{c|c}
    \hline
    Element & Width ($1$/Bohr$^2$) \\
    \hline
     C  &  22.5 \\
     H  &  4.5\\
     O  &  13 \\
     \hline
    \end{tabular}
    \caption{Gaussian widths used for the various atom types in the AIMS TBFs.}
    \label{tab:gaussian_widths}
\end{table}

\bibliography{references}